%% file: main.tex
\documentclass[11pt,a4paper]{elsarticle}

\UseRawInputEncoding
\pdfoutput=1
\makeatletter
\def\ps@pprintTitle{%
 \let\@oddhead\@empty
 \let\@evenhead\@empty
 \def\@oddfoot{\centerline{\thepage}}%
 \let\@evenfoot\@oddfoot}
\makeatother

\usepackage{verbatim}
\usepackage[UKenglish]{babel}
\usepackage[UKenglish]{isodate}
\cleanlookdateon
\usepackage{epsfig}
\usepackage{graphicx}
\graphicspath{ {figs/} }
\usepackage[margin=15pt,font={footnotesize},labelfont=bf,format=hang]{caption}
\usepackage[margin=15pt,font={footnotesize},labelfont=bf,format=hang]{subcaption}
\usepackage{bm}
\usepackage{amsbsy}
\usepackage{amssymb}
\usepackage{amsmath}
\usepackage{mathtools}
\usepackage{isomath}
\usepackage{enumitem}
\usepackage[pdfencoding=auto]{hyperref}
\usepackage[dvipsnames]{xcolor}
\usepackage{algorithm}
\usepackage{algorithmicx}
\usepackage{xpatch}
\usepackage[normalem]{ulem}
\usepackage{booktabs}
\usepackage{soul}
\usepackage{siunitx}
\usepackage{color}
\usepackage{tikz}
\usepackage[section]{placeins}
\usepackage[margin=2.5cm]{geometry}

\renewcommand{\v}[1]{\mathbfit{#1}}      
\newcommand{\uv}[1]{\widehat{\mathbfit{#1}}}      
\newcommand{\q}[1]{\mathsfbfit{#1}}        
\renewcommand{\t}[1]{\mathsfbfit{#1}}	 
\newcommand{\vu}[1]{\mathbf{#1}}         
\newcommand{\tu}[1]{\bm{\mathsf{#1}}}	 

\newcommand{\qprod}{\bullet} 

\renewcommand{\d}{{\mathrm d}}
\def \p {\partial}

\newcommand{\vdel}{\boldsymbol{\nabla}}

\newcommand{\vnabla}{\bm{\nabla}}

\renewcommand{\hat}{\widehat}
\renewcommand{\tilde}{\widetilde}

\newcommand{\fd}[2]{\mathchoice{\frac{\d #1}{\d #2}}{\d #1/\d #2}{\d #1/\d #2}{\d #1/\d #2}}

\newcommand{\pd}[2]{\mathchoice{\frac{\p #1}{\p #2}}{\p #1/\p #2}{\p #1/\p #2}{\p #1/\p #2}}

\newcommand{\order}{\mathcal{O}}

\newcommand{\Reals}{\mathbb{R}}

\newcommand{\correction}[1]{{\color{black}#1}}

\usepackage{cleveref}
\crefformat{appendix}{#2#1#3} 

\journal{Journal of Computational Physics}
\begin{document}

\begin{frontmatter}

\title{A generalised drift-correcting time integration scheme for Brownian suspensions of rigid particles with arbitrary shape}

\author[mymainaddress]{Timothy A Westwood}

\author[mysecondaryaddress]{Blaise Delmotte}

\author[mymainaddress]{Eric E Keaveny}

\address[mymainaddress]{Department of Mathematics, Imperial College London, South Kensington Campus, London, SW7 2AZ, UK}
\address[mysecondaryaddress]{LadHyX, CNRS, Ecole Polytechnique, Institut Polytechnique de Paris, 91120 Palaiseau, France}

\begin{abstract}
The efficient computation of the overdamped, random motion of micron and nanometre scale particles in a viscous fluid requires novel methods to obtain the hydrodynamic interactions, random displacements and Brownian drift at minimal cost.  Capturing Brownian drift is done most efficiently through a judiciously constructed time-integration scheme that automatically accounts for its contribution to particle motion.  In this paper, we present a generalised drift-correcting (gDC) scheme that accounts for Brownian drift for suspensions of rigid particles with arbitrary shape.  The scheme seamlessly integrates with fast methods for computing the hydrodynamic interactions and random increments and requires a single full mobility solve per time-step.  As a result, the gDC provides increased computational efficiency when used in conjunction with grid-based methods that employ fluctuating hydrodynamics to obtain the random increments.  Further, for these methods the additional computations that the scheme requires occur at the level of individual particles, and hence lend themselves naturally to parallel computation.  We perform a series of simulations that demonstrate the gDC obtains similar levels of accuracy as compared with the existing state-of-the-art.  In addition, these simulations illustrate the gDC's applicability to a wide array of relevant problems involving Brownian suspensions of non-spherical particles, such as the structure of liquid crystals and the rheology of complex fluids.
\end{abstract}

\begin{keyword}
Brownian motion, suspensions, time integration, fluctuations, simulation
\end{keyword}

\end{frontmatter}
\tableofcontents

\section{Introduction}

Brownian motion, the random motion exhibited by colloidal particles due to thermal fluctuations in the surrounding viscous fluid, is a common and important feature of fluidic systems at the micron and nanometre scales \cite{Russel1981,russel1991colloidal,graham2018microhydrodynamics}.  It plays a fundamental role in determining the distribution and configuration of polymers and particles that comprise complex fluids \cite{doi_theory_1986,larson1999structure}, which then in turn affect suspension rheology and bulk mechanical response to applied stresses \cite{Batchelor1977,Bossis1989,Foss2000}.  Further, Brownian motion and thermal fluctuations play a critical role in biological and cellular transport processes, often in conjunction with the cell's active mechanisms that utilise chemical energy \cite{bressloff2013stochastic}.

To simulate these systems, one often considers the equations of motion in the Brownian dynamics \cite{ermak_brownian_1978}, or overdamped \cite{pavliotis2014stochastic}, limit where momentum variables are taken to have reached thermal equilibrium and the velocity field in the surrounding fluid is governed by the steady Stokes equations.  In the context of rigid particles, this means that the dynamics of the particle positions and orientations are provided by the overdamped Langevin equations, and accordingly, the dynamics of the corresponding probability distribution of the particle positions and orientations is described by Smoluchowski's equation.  

The hydrodynamic interactions between the particles due to the surrounding fluid are embodied in the dense, position-dependent mobility matrix \cite{Kim1991,Happel2012} that relates the generalised forces on the particles to the resulting particle velocities.  The hydrodynamic interactions between the particles also impact the fluctuations that they experience.  The fluctuation-dissipation theorem \cite{kubo_fluctuation-dissipation_1966} stipulates that the covariance of the random particle velocities is proportional to the mobility matrix and hence so too is the particle diffusion matrix.  This requires the random particle velocities to be proportional to the square root of the mobility matrix.  Finally, due to hydrodynamic interactions, the overdamped limit produces a nontrivial thermal, or Brownian, drift term \cite{ermak_brownian_1978} in the equations of motion that is proportional to the divergence of the mobility matrix with respect to the particle positions and orientations.  

Initially, the computations of these three quantities -- the mobility matrix, random velocities, and Brownian drift -- in approaches such as Stokesian Dynamics \cite{Brady1988} utilised a variety of direct methods.  The entries of the mobility matrix were determined pairwise from analytical expressions derived in relevant asymptotic limits and the random particle velocities were computed using a Cholesky decomposition of the resulting matrix.  Finally, Brownian drift was incorporated directly by evaluating expressions for the derivatives of the mobility matrix entries.  A reliance on these direct approaches limited system sizes, as well as particle shapes to spheres.  

Since these first computations, there have been several key advances, some of which have been incorporated into Stokesian dynamics \cite{sierou2001accelerated,Banchio2003}, including fast, matrix-free methods for determining the particle velocities given the forces applied to them.  These methods provide the action of the mobility matrix on a vector in $O(N \log N)$ operations, where $N$ is the vector length, without ever computing the matrix itself.  Such approaches are built around fast summation techniques such as the FMM \cite{greengard1987fast,tornberg2008fast,ying2004kernel}, or the FFT and include spectrally-accurate Ewald summation \cite{lindbo2011spectral}, positively split Ewald summation \cite{Fiore2017,Fiore2018}, Accelerated and Fast Stokesian Dynamics \cite{sierou2001accelerated,Banchio2003,wang2016spectral,fiore2019fast}, in addition to the immersed boundary \cite{Peskin2002} or similar methods, such as the force-coupling method \cite{maxey_localized_2001-1,lomholt_force-coupling_2003-1}, that utilise fast, grid-based solvers.  

Recent work on computational methods for Brownian suspensions has focused on developing similarly rapid computations of the random particle velocities that can be used with the matrix-free methods described above.  While the early work of Fixman \cite{Fixman1986} relying on a Chebyshev expansion of the spectrum of the mobility matrix has been adopted in implementations of Accelerated Stokesian Dynamics \cite{Banchio2003} and Brownian Dynamics \cite{jendrejack2000hydrodynamic}, only more recently have techniques been developed that can be used more seamlessly with fast summation methods.  These include the Lanczos algorithm \cite{chow_preconditioned_2014} which computes iteratively an approximation of the matrix-square root using the Ritz values and vectors generated at the final iteration.  For methods that utilise grid-based solvers, the random particle velocities can be computed rapidly from the flows generated by the spatially uncorrelated fluctuating stress originally considered by Landau and Lifshitz \cite{Landau1959}.  This fluctuating hydrodynamics approach provides the foundation for several methods that capture particle Brownian motion including Lattice Boltzmann \cite{Ladd1993}, distributed Lagrange multiplier technique \cite{Sharma2004}, finite element methods \cite{de2016finite}, the fluctuating \cite{Delong2014} and stochastic \cite{atzberger2007stochastic,kramer2008foundations} immersed boundary methods and the fluctuating force-coupling method \cite{keaveny_fluctuating_2014-1,delmotte_simulating_2015}. Further, computations that rely on decomposing the mobility matrix, as is the case with positively split Ewald summation \cite{Fiore2017,Fiore2018} or discretisations of the fluctuating boundary integral equations \cite{bao2018fluctuating}, may use the fluctuating hydrodynamics approach and the Lanczos algorithm side-by-side to compute different contributions to the particle stochastic velocities resulting from the decomposition.

The final aspect of the computation is to account for Brownian drift when advancing the particle positions.  This is done most efficiently using a well-designed time integration scheme that automatically accounts for Brownian drift, yielding a numerical solution whose moments converge to the their true values as the time step size goes to zero.  Fixman again provides \cite{Fixman1978,Grassia1995} an early example of a midpoint scheme which, to leading order in the time step size, emits the Brownian drift term as part of its error expansion.  This scheme, however, relies on a resistance formulation, where the dense, long-ranged mobility matrix must be inverted, introducing costly additional linear systems.  Recent work, however, has shown that similar integration schemes can be constructed that avoid these additional linear systems.  It has been shown \cite{Delong2014} that the drift can be incorporated into the Euler-Maruyama scheme using random finite differences (RFD).  This involves applying the mobility matrix evaluated at randomly displaced positions to a random force vector with an appropriate covariance.  This limits additional costs due to mobility matrix multiplications.  For the grid-based methods built around fluctuating hydrodynamics, the drifter-corrector (DC) midpoint scheme \cite{delmotte_simulating_2015} eliminates the additional mobility matrix multiplication completely.  Here, the drift term is incorporated by advancing the particle positions to the midpoint using the particle velocity extracted from the flow generated by the fluctuating stress.  The full mobility computation is then performed at the midpoint.  

While RFD and the DC represent important advances in designing schemes that capture Brownian drift and naturally interface with matrix-free methods, they were originally limited to the case of spherical particles with mobility matrices obtained through singular multipole expansions, or their regularised equivalents.  Recent work has extended the RFD approach for simulation of rigid particles of arbitrary shape that are represented by multiple discrete degrees of freedom constrained to move as a rigid body.  Sprinkle \textit{et al.} \cite{sprinkle_large_2017,Sprinkle2019} proposed a family of time integration schemes that use the chain rule to split the computation of the divergence of the body mobility matrix into three contributions that can be obtained at a lower cost using RFD. The resulting schemes, called Euler-Maruyama Traction (EM-T) and Trapezoidal Slip (T-S), are both weakly first-order  accurate, but achieve first (resp.\ second) order accuracy for deterministic problems and require two (resp.\ three) full mobility solves per time step.

In this work, we generalise the DC scheme to simulate rigid bodies of arbitrary shape.  Like the original DC, the generalised DC (gDC) scheme requires a single full mobility solve per time step.  This significantly accelerates time integration for schemes built around grid-based solvers that can take advantage of the flows generated by a fluctuating stress.  The gDC may also be used for matrix-based computations at a cost comparable to the existing state-of-the-art.  The main idea behind the scheme is to advance to the mid-step using particle velocities obtained by orthogonally projecting the fluctuating velocities onto the space of rigid body motions.  At the midstep, the full mobility is treated and important factors based on the divergence of the projected fluctuating velocity are incorporated into the update.  The gDC has the nice property that many of the additional computations needed to capture Brownian drift occur at the level of individual particles, and hence naturally lend themselves to parallel computation and larger-scale simulation.  We show that the resulting scheme is weakly first-order accurate and, in practice, provides errors similar in magnitude to the T-S scheme.  Finally, we demonstrate the applicability of the scheme for larger-scale simulation by considering confined suspensions of rod-like particles, as well as the rheology of Czech hedgehog particle suspensions.  

\section{Brownian Dynamics of rigid particles of arbitrary shape}

\begin{figure}[htb!]
\centering
\def\svgwidth{0.9\columnwidth}
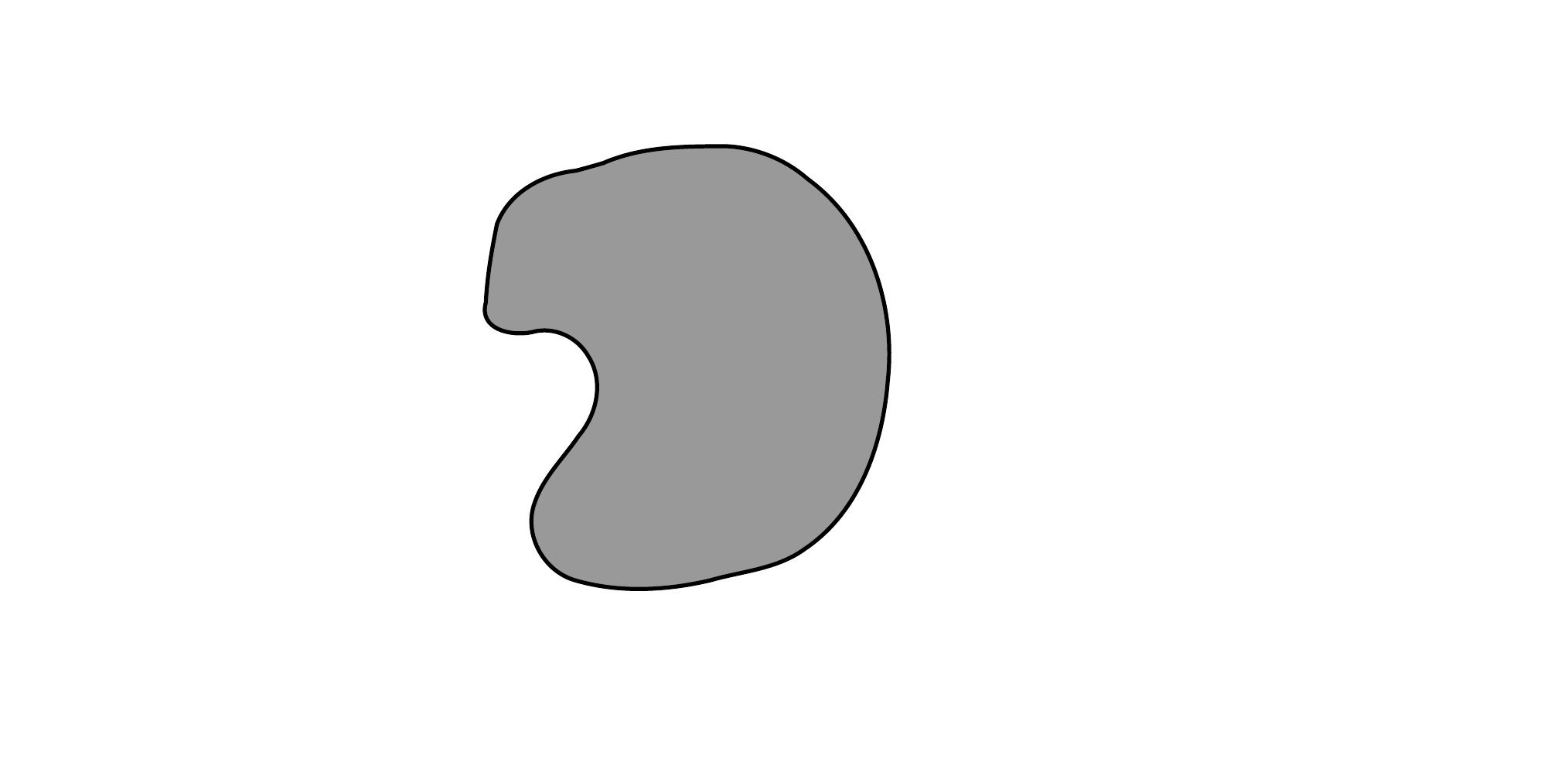
\caption{Illustration of a rigid body with position $\v{Y}_p$ and orientation $\v{q}_p$.}
\label{fig:rigidbodysketch}
\end{figure}

Consider a suspension of $N$ rigid particles where the centre of mass position of particle $p$ at time $t$ is $\v{Y}_p(t)$, while the rotation relative to its initial orientation is given by the unit quaternion $\q{q}_p(t)$ (see \cref{fig:rigidbodysketch}).  We provide a detailed overview of using quaternions to represent rotations in \cref{sec:quaternions}, but present some of the key facts here.  Any vector $\v{b}$ in the body frame of particle $p$ is given by $\v{B}(t) = \t{R}(\q{q}_p(t))\v{b}$ in the lab frame at time $t$.  The rotation matrix $\t{R}$ is related to entries of the quaternion through
\begin{align}
\label{equation:quaternion_rotation_matrix}
    \t{R}\left(\q{q}\right) = \left[\begin{matrix} 1 - 2q^{2}_{2} - 2q^{2}_{3} & 2\left(q_{1}q_{2} - q_{3}q_{0}\right) & 2\left(q_{1}q_{3} + q_{2}q_{0}\right) \\
2\left(q_{1}q_{2} + q_{3}q_{0}\right) & 1 - 2q^{2}_{1} - 2q^{2}_{3} & 2\left(q_{3}q_{2} - q_{1}q_{0}\right) \\
2\left(q_{1}q_{3} - q_{2}q_{0}\right) & 2\left(q_{3}q_{2} + q_{1}q_{0}\right) & 1 - 2q^{2}_{2} - 2q^{2}_{1}
\end{matrix}\right].    
\end{align}
The quaternions are advanced in time using the formula
\begin{align}
    \q{q}_p(t) = \exp\left(\v{u}_p(t)\right)\qprod\q{q}_p(0),
\end{align}
where $\qprod$ is the Hamiltonian product (see \cref{sec:quaternions}), $\v{u} \in \Reals^3$ is the Lie algebra element, and 
\begin{align}
\exp\left(\v{u}\right) = \left(\cos\left(\frac{\|\v{u}\|}{2}\right),\;\sin\left(\frac{\|\v{u}\|}{2}\right)\frac{\v{u}}{\|\v{u}\|}\right).
\end{align}
We will be interested in numerically integrating the overdamped Langevin equations for hydrodynamically interacting rigid bodies that govern the positions, $\v{Y}_p$, and Lie algebra elements, $\v{u}_p$, for each rigid body $p$. 

\subsection{The overdamped Langevin equation for rigid body motion}

Suppose that the particles are subject to an external potential $U(\v{Y}_1, \v{u}_1, \hdots, \v{Y}_N, \v{u}_N)$.  The force on particle $p$ is then $-\p_{\v{Y}_p}U$, while the corresponding torque is $-\t{D}_{\v{u}_p}^{-\top}\p_{\v{u}_p}U$.  A derivation of this expression for the torque is presented in \cref{sec:lie_torque}.  The matrix $\t{D}_{\v{u}_p}^{-1}$ is the $3\times 3$ `dexpinv' matrix defined by
\begin{align} \label{eqn:dexpinv}
        \t{D}^{-1}_{\v{u}_p} = \t{I} - \frac{1}{2}\left[\v{u}_p\times\right] - \frac{1}{2\|\v{u}_p\|^2}\left(\|\v{u}_p\|\cot\left(\frac{\|\v{u}_p\|}{2}\right) - 2\right)\left[\v{u}_p\times\right]^2,
\end{align}
where $\left[\v{v}\times\right]_{ij} = \varepsilon_{ikj}v_k$ and $\left[\times\v{v}\right] = -\left[\v{v}\times\right] = \left[\v{v}\times\right]^\top$ for any $\v{v} \in \Reals^3$.  The `dexpinv' matrix relates the angular velocity of particle $p$, $\v{\Omega}_p$, to the time derivative of $\v{u}_p$ through
\begin{align}
    \fd{\v{u}_p}{t} = \t{D}^{-1}_{\v{u}_p}\v{\Omega}_p.
\end{align}

In the overdamped limit, the force and torque vectors are linearly related to their contributions to the particle velocity and angular velocity vectors via the $6N\times 6N$ mobility matrix, $\t{N}$, whose entries depend on the particle configuration. The application of the mobility matrix is equivalent to solving the Stokes boundary value problem for a collection of rigid particles subject to applied forces and torques.  Thus, in the absence of Brownian motion, the equations of motion are given by the system of differential equations
\begin{align}
\fd{\v{x}}{t} = -\t{\tilde{N}}\p_\v{x}U,
\end{align}
where $\v{x} = \left[\v{Y}_1^\top, \v{u}_1^\top, \hdots, \v{Y}_N^\top, \v{u}_N^\top\right]^\top$ describes the positions and orientations of all particles and $\t{\tilde{N}} = \t{\Phi}\t{N}\t{\Phi}^\top$ with
\begin{align}
    \t{\Phi} = \left[\begin{matrix}\t{I}_3 & \tu{0} & \hdots & \hdots & \tu{0} \\ \tu{0} & \t{D}^{-1}_{\v{u}_1} & & & \vdots \\ \vdots & & \ddots & & \vdots \\ \vdots & & & \t{I}_3 & \tu{0} \\ \tu{0} & \hdots & \hdots & \tu{0} &\t{D}^{-1}_{\v{u}_N} \end{matrix}\right].
\end{align}

When Brownian motion is present, the particles will also move randomly due to thermal fluctuations in the surrounding fluid. In the overdamped limit, the effects of Brownian motion are captured through the inclusion of random increments of the positions and Lie algebra elements, turning the equations of motion into a system of stochastic differential equations. Specifically, these increments are given by $\sqrt{2k_BT}\t{\tilde{N}}^{1/2}d\v{W}$, where $k_BT$ is the thermal energy, $\tilde{\t{N}}^{1/2} = \t{\Phi}\t{N}^{1/2}$ with $\t{N}^{1/2}$ being the matrix square root of the mobility matrix, and $\v{W}$ is a $6N \times 1$ vector of independent Wiener processes. The dependence on the mobility matrix ensures that the fluctuation-dissipation theorem is satisfied, a necessary condition for the Boltzmann distribution to be recovered at equilibrium. 

Finally, along with the random velocities, the overdamped equations of motion require the inclusion of the thermal drift term, $k_BT\partial_\v{x}\cdot\tilde{\t{N}}dt$ \citep{ermak_brownian_1978}. This ensures that the stochastic differential equation with an It\^o interpretation of the stochastic integral yields dynamics consistent with those of Smoluchowski's equation for the corresponding probability distribution.  Putting these terms together, we arrive at the equations of motion for the particle positions and orientations,
\begin{equation} 
    d\v{x} = \left(-\tilde{\t{N}}\p_\v{x}U + k_BT\partial_\v{x}\cdot\tilde{\t{N}}\right)dt + \sqrt{2k_BT}\tilde{\t{N}}^{1/2}d\v{W}.
    \label{eqn:system_sde}
\end{equation}

\subsection{Mobility matrix} \label{sec:mobmat}

\begin{figure}[htb!]
\centering
\def\svgwidth{0.3\columnwidth}
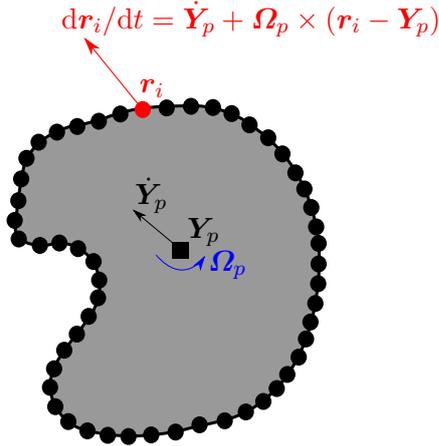
\caption{Discretisation of the rigid body surface with discrete degrees of freedom.}
\label{fig:discretisedbody}
\end{figure}

The purpose of this paper is to develop time integration schemes for \cref{eqn:system_sde} that avoid the direct computation of the thermal drift term and naturally interface with matrix-free methods for computing the random increments and particle velocities arising from the applied forces.  In doing so, we will assume that the particles are discretised into $M$ total degrees of freedom, as depicted in \cref{fig:discretisedbody}.  This discretisation can involve the surface of the particles, as is done when considering the first-kind boundary integral representation of the Stokes equations \cite{Pozrikidis,Kim1991}, or can be related to a volume discretisation where elements of the particle volume are represented by regularised distributions of force, for example. 

Let $\v{r}_i$ denote the position of discrete degree of freedom $i$, $\v{v}_i \equiv d\v{r}_i/dt$ be its velocity and $\v{\lambda}_i$ be the force it exerts on the fluid.  The hydrodynamic interactions between the discrete degrees of freedom provide a linear relationship between $\v{v}_i$ and $\v{\lambda}_i$ such that
\begin{equation}
 \v{v}_i = \sum_{j=1}^{M}\t{M}_{ij}\v{\lambda}_j,
 \label{eq:ddof_mob}
\end{equation}
where $\t{M}_{ij}$ is the matrix that relates the force on discrete degree of freedom $j$ to the velocity of discrete degree of freedom $i$.  Additionally, the requirement that all discrete degrees of freedom belonging to particle $p$ move as a single rigid body (see \cref{fig:discretisedbody}) stipulates that
\begin{equation}
 \v{v}_i = \dot{\v{Y}}_{p} + \v{\Omega}_p\times\left(\v{r}_i - \v{Y}_p\right),\,\, \forall i\in \mathcal{B}_p,
 \label{eq:rigid_body_cond}
\end{equation}
where $\dot{\v{Y}}_{p}$ is the translational velocity of particle $p$, $\v{\Omega}_p$ is its angular velocity and $\mathcal{B}_p$ is the set of discrete degrees of freedom belonging to particle $p$. 

Introducing the vectors $\v{\lambda} = [\v{\lambda}_1^{\top},\hdots, \v{\lambda}_M^{\top}]^{\top}$ and $\v{V} = [\v{V}_1^{\top}, \dots, \v{V}_{N}^{\top}]^{\top}$ with $\v{V}_p =[\dot{\v{Y}}_{p}^{\top}, \v{\Omega}_p^{\top}]^{\top}$, \cref{eq:ddof_mob,eq:rigid_body_cond} for all particles $p = 1,\dots,N$ can be equated to yield
\begin{equation}
  \t{M}\v{\lambda} = \t{K}\v{V},
 \label{eq:rigid_body_cond_short}
\end{equation}
where 
\begin{align}
  \t{M} = \left[\begin{array}{ccc}
\t{M}_{11} & \dots & \t{M}_{1M}\\
\vdots & \dots & \vdots\\
\t{M}_{M1} & \dots & \t{M}_{MM}\\
\end{array}\right]
\end{align}
is the $3M\times 3M$ mobility matrix for the discrete degrees of freedom and $\t{K}$ is a $3M\times 6N$ block matrix that maps the velocity and angular velocity from the space of rigid body motions to the velocities of the discrete degrees of freedom.  Specifically 
\begin{align}
    \t{K}_{i}^{(p)} = \begin{cases}
\left[\begin{array}{cc}
\t{I}_3 & \left[\times (\v{r}_i - \v{Y}_p)\right]\end{array}\right],  &  \mbox{if } i\in \mathcal{B}_p,\\
\\
\left[\begin{array}{cc}
\vu{0}_3 & \vu{0}_3\end{array}\right], & \mbox{otherwise,} \\
\end{cases} 
\end{align}
is the $3\times6$ block that relates the rigid body motion of particle $p$ to the translational velocity of discrete degree of freedom $i$.  Note that the specific form of $\t{M}$ will depend on how the discretisation is performed, however, here we will assume that the discretisation results in a positive definite $\t{M}$, thereby preserving the structure of the underlying continuous equations.  This is an important ingredient for the fluctuation-dissipation theorem to be satisfied exactly in the discrete setting.

In the absence of inertia, the conservation of linear and angular momentum reduce to the balance of force and torque, which can be written compactly for the $N$ particles as
\begin{equation}
  \t{K}^\top\v{\lambda} = \v{F},
 \label{eq:force_torque_balance_compact}
\end{equation}
where $\v{F}$ is the total force and torque vector applied on the particles.

The mobility problem given by \cref{eq:rigid_body_cond_short,eq:force_torque_balance_compact} forms a saddle point system 
\begin{align}
    \left[\begin{array}{cc}
\t{M} & -\t{K}\\
-\t{K}^\top & \vu{0}\\
\end{array}\right]\left[\begin{array}{c}
\v{\lambda}\\
\v{V}\\
\end{array}\right]=\left[\begin{array}{c}
0 \\
-\v{F}\\
\end{array}\right].
\label{eq:saddle_point_sys}
\end{align}
After eliminating $\v{\lambda}$, one obtains 
\begin{align}
\v{V} = \left(\t{K}^\top \t{M}^{-1} \t{K}\right)^{-1}\v{F}
\end{align}
and therefore the mobility matrix is given by
\begin{align}
 \t{N} = \left(\t{K}^\top \t{M}^{-1} \t{K}\right)^{-1}.
 \label{eq:body_mob}
\end{align}

As we describe later, our discretisation will follow from the rigid blob framework \cite{Balboa2017}, but the gDC time integration scheme is applicable to any spatial discretisation provided that the resulting mobility matrix $\t{N}$ can be related to the saddle point system in \cref{eq:saddle_point_sys}. 
This particular decomposition has the advantage that the computation of the stochastic particle velocities can be readily incorporated into the system by including a random velocity for the discrete degrees of freedom \cite{sprinkle_large_2017}.  Specifically, we consider the system
\begin{align}
    \left[\begin{array}{cc}
\t{M} & -\t{K}\\
-\t{K}^\top & \vu{0}\\
\end{array}\right]\left[\begin{array}{c}
\v{\lambda}\\
\v{V}\\
\end{array}\right]=\left[\begin{array}{c}
-\breve{\v{v}}\\
-\v{F}\\
\end{array}\right].
\label{eq:lin_sys_RMB}
\end{align}
with the random velocity,
\begin{align}
 \breve{\v{v}} = \sqrt{2k_BT}\t{M}^{1/2}\v{W}(t),
 \label{eq:v_brown_RMB}
\end{align}
where $\t{M}^{1/2}$ is the square root of $\t{M}$ which satisfies $\t{M}^{1/2}(\t{M}^{1/2})^\top =\t{M}$, and $\v{W}(t)$ is a $3M\times 1$ vector of Wiener processes.
After eliminating $\v{\lambda}$, one obtains 
\begin{align}
\v{V} = \t{N}\v{F} +  \breve{\v{V}}
\end{align}
where $\breve{\v{V}} = \t{N}\t{K}^\top\t{M}^{-1}\breve{\v{v}}$.  As a result the fluctuating body velocities are given by 
\begin{align}
 \breve{\v{V}} &= \sqrt{2k_BT}\t{N}\t{K}^\top\t{M}^{-1}\t{M}^{1/2}\v{W}(t) = \sqrt{2k_BT}\t{N}^{1/2}\v{W}(t),
\end{align}
where $\t{N}^{1/2}$ satisfies the fluctuation dissipation balance with $\t{N}^{1/2}(\t{N}^{1/2})^\top = \t{N}$.

As shown by Balboa-Usabiaga \textit{et al.} \cite{usabiaga_hydrodynamics_2016}, the symmetric saddle point problem \eqref{eq:lin_sys_RMB} can be efficiently solved with an iterative Krylov subspace method, such as GMRES, using a simple block diagonal preconditioner $\t{P}$. The matrix $\t{P}$ is constructed by setting to zero the entries of $\t{M}$ in \eqref{eq:lin_sys_RMB} if they correspond to interactions between discrete degrees of freedom from different rigid bodies.  As $\t{K}$ is a block diagonal matrix, the cost of solving the linear system \eqref{eq:lin_sys_RMB} will be related primarily to the costs of applying the matrix $\t{M}$ and computing $\breve{\v{v}}$.  

\section{The generalised drifter-corrector (gDC)}
\label{sec:gDC_alg}
The primary result of this paper is the following generalisation of the DC scheme to rigid bodies of arbitrary shape.  The gDC updates the particle positions and Lie algebra elements while automatically accounting for the Brownian drift term that arises in the equations of motion.  It has the advantage of requiring only a single full mobility solve per time step.  For grid-based solvers, the gDC incurs nearly the same cost as applying the well-known Euler-Maruyama scheme, outlined in \cref{sec:EM_alg}, which does not capture the Brownian drift.  For matrix-based approaches the cost will be similar to the existing state-of-the-art \cite{sprinkle_large_2017,Sprinkle2019}.

Below we present the gDC algorithm. The time-step is indicated by superscript, $n$, e.g. $\v{\mathcal{V}}^n$ would correspond to $\v{\mathcal{V}}\left(\v{x}^n\right) = \v{\mathcal{V}}\left(\v{x}(t^n)\right)$ in the continuous setting. For the purposes of implementing the algorithm, it should be noted that the collection of Lie algebra elements contained in $\v{x}^n$ is set to zero at the start of each time step (see \cref{sec:quaternions}).\\

\noindent For each time step $n = 0,1,2,...$
\begin{enumerate}

    \item Generate a vector  $\v{W}$ (or tensor $\t{W}$ if using fluctuating hydrodynamics) of $\mathcal{N}(0,1)$ random variables.
    
    \item Solve the following saddle point problem at the start of the time-step,
    \begin{equation}
        \begin{bmatrix}
        \t{I} & -\t{K}^n \\
        -\left(\t{K}^\top\right)^n  & \vu{0}
        \end{bmatrix}
        \begin{bmatrix}
        \v{\lambda}^n \\ \v{\mathcal{V}}^n
        \end{bmatrix}
         = \begin{bmatrix}
         -\breve{\v{v}}^n = -\sqrt{\frac{2 k_B T}{\Delta t}}\left(\t{M}^{1/2}\right)^n\v{W} \\ \vu{0}
         \end{bmatrix},
         \label{eq:unconst}
    \end{equation}
    yielding
    \begin{equation}
    \v{\mathcal{V}}^n = \left(\left(\t{K}^\top\right)^n \t{K}^n\right)^{-1}\left(\t{K}^\top\right)^n \breve{\v{v}}^n.
    \label{eq:sol_unconst}
    \end{equation}
    Note that this is also the solution of the least-squares problem $\min_{\v{\mathcal{V}}^n}\|\t{K}^n\v{\mathcal{V}}^n - \breve{\v{v}}^n\|$.
    
    \item Define $\nu = 1 + \frac{\Delta t}{2}\left(\partial_\v{x}\cdot\v{\mathcal{V}}\right)^n$. For a small parameter $\delta$, the divergence can be calculated via\footnote{In order to respect the physical units of the problem and to minimize the variance of the RFD approximation, we multiply  $\delta$ by a typical length scale $L_p$  when computing the translational part of $\left(\p_\v{x}\cdot\v{\mathcal{\mathcal{V}}}\right)^n$ \cite{sprinkle_large_2017}. $L_p$ is the maximum distance between two discrete degrees of freedom in the body and represents the typical size of body $p$.} 
    \begin{enumerate}\item finite-differencing:
    \begin{align}
     \left(\partial_\v{x}\cdot\v{\mathcal{V}}\right)^n = \sum_{i=1}^6\sum_{p=1}^N \frac{\mathcal{V}^{(p)}_i\left(\v{x}^n+\delta\v{e}^{(p)}_i\right)-\mathcal{V}^{(p)}_i\left(\v{x}^n\right)}{\delta} + \order(\delta^2)
     \label{eq:FD}
    \end{align}
    where $\mathcal{V}^{(p)}_i$ is the $i$-th component of the velocity of particle $p$, and  $\v{e}^{(p)}_i$  is a $6N\times 1$ vector whose nonzero entries, corresponding to the position and orientation of particle $p$, are given by the  $i$-th Cartesian basis vector of $\mathbb{R}^6$; or \item random finite-differences:
    \begin{equation}\label{eqn:rfd_div}
        \left(\p_\v{x}\cdot\v{\mathcal{\mathcal{V}}}\right)^n = \Biggl\langle\sum_{i=1}^{6}\sum_{p=1}^N\frac{\tilde{\v{W}}\cdot\v{e}^{(p)}_i}{\delta}\left(\mathcal{\mathcal{V}}^{(p)}_i(\v{x}^n + \delta\tilde{\v{W}}) - \mathcal{\mathcal{V}}^{(p)}_i(\v{x}^n)\right)\Biggr\rangle + \order(\delta^2),
    \end{equation}
    where $\tilde{\v{W}}$ is a vector of independent $\mathcal{N}(0,1)$ random variables which is also independent of $\v{W}$. By replacing the divergence in the definition of $\nu$ with the term inside the expectation in \cref{eqn:rfd_div}, the appropriate value of $\nu$ will be achieved in expectation, which is ultimately all that is required to demonstrate weak-accuracy. 
    \end{enumerate}

    We find that option (a) is best partnered with grid-based mobility methods, whilst option (b) is best suited to matrix-based methods to reduce the computational cost. We discuss this in more detail in \cref{sec:gDC_cost}.
    
    \item Move the body positions and orientations to the mid-step according to
    \begin{equation}
        \v{x}^m = \v{x}^n + \frac{\Delta t}{2}\v{\mathcal{V}}^n.
    \end{equation}

    \item Solve the full mobility problem at the mid-step,
    \begin{equation}
        \begin{bmatrix}
        \t{M}^m & -\t{K}^m \\
        -\left(\t{K}^\top\right)^m  & \vu{0}
        \end{bmatrix}
        \begin{bmatrix}
        \v{\lambda}^m \\ \v{V}^m
        \end{bmatrix}
         = \begin{bmatrix}
         -\breve{\v{v}}^m = -\sqrt{\frac{2 k_B T}{\Delta t}}\left(\t{M}^{1/2}\right)^m\v{W} \\ -\v{F}^m
         \end{bmatrix},
        \label{eq:step5}
    \end{equation}
    for
    \begin{equation}
    \v{V}^m = \t{N}^m\left(\v{F}^m + \left(\t{K}^\top\right)^m\left(\t{M}^{-1}\right)^m\breve{\v{v}}^m\right) = \t{N}^m\v{F}^m + \breve{\v{V}}^m ,
    \end{equation}
    where $\v{F}^m$ contains the forces and torques acting on the bodies at the mid-step. Note that quantities are evaluated at the mid-step using the mid-step quaternions $\q{q}_p^m = \exp\left(\v{u}_p^m\right)\qprod\q{q}_p^n$.
    
    \item Update the positions and orientations of the bodies according to
    \begin{align}
        \v{x}^{n+1} &= \v{x}^n + \nu\Delta t\v{V}^m,
    \end{align}
    with the new orientation quaternion of each particle $p$ defined by $\q{q}^{n+1}_p = \exp\left(\v{u}^{n+1}_p\right)\qprod\q{q}_p^n$.
\end{enumerate}

In words, the gDC first computes a set of particle velocities by solving a least squares problem to find the appropriate stochastic velocity, $\v{\mathcal{V}}$, in the space of rigid body motions. This velocity is used to advance the particle positions to the mid-step where the full computation is performed. The positions are then updated using the velocity computed at the mid-step scaled by a factor based on the divergence of  $\v{\mathcal{V}}$. As shown in \cref{sec:consistency}, the scheme recovers the first and second moments of the increment $\Delta \v{x} = \v{x}^{n+1}-\v{x}^{n}$ to first-order in expectation.

\section{Applying $\t{M}$ and computing $\breve{\v{v}}$}
\label{sec:HI}

As shown in \cref{sec:mobmat}, hydrodynamic interactions between bodies are directly obtained from the matrix $\t{M}$.  The action of $\t{M}$ on a vector is required to compute the deterministic velocities $\v{V} = \t{N}\v{F}$, while the product $\t{M}^{1/2} \v{W}$ is necessary to obtain the Brownian velocities $\breve{\v{V}}$. It is therefore essential to use and develop efficient methods to compute the action of $\t{M}$ and its square root.

In this section, we outline two distinct strategies with similar resolution of hydrodynamic interactions between the discrete degrees of freedom. The first follows directly from an implementation of the rigid multiblob model \cite{Balboa2017}  where the rigid particles are discretised into surface or volume elements that interact via the Rotne-Prager-Yamakawa (RPY) tensor.  The other approach called the fluctuating Force Coupling Method (FCM) is matrix-free and simultaneously applies $\t{M}$ and computes $\breve{\v{V}}$ by solving the fluctuating Stokes equations on a grid with a forcing term that accounts for the presence of the particles.  

\subsection{RPY tensor}
\label{sec:RPY}
The well-known RPY mobility matrix was originally developed to provide a symmetric positive definite, pairwise approximation of the mobility matrix for a collection of spherical particles of equal radii in an unbounded domain \cite{wajnryb_generalization_2013}.  Extensions of the RPY matrix for particles of different radii \cite{zuk_rotneprageryamakawa_2014}, in a background shear flow \cite{wajnryb_generalization_2013}, and above a no-slip boundary \cite{swan_simulation_2007} are available in the literature. Following the rigid multiblob model, the RPY tensor can be used to provide the hydrodynamic interactions between the discrete degrees of freedom making up the rigid particles.  When discretising a particle surface, the rigid multiblob model provides a first-order accurate approximation to the particle mobility.  

In general, the computational cost of these matrix-vector products scales quadratically with the number of discrete degrees of freedom. More sophisticated methods, such as the fast multipole methods (FMM) \cite{Liang2013} and Ewald methods, achieve a linear scaling.  For periodic domains, one can use the positively split Ewald method \cite{Fiore2017,Fiore2018} to compute the action of $\t{M}$ on a vector.  In our computations below, we perform a direct pairwise evaluation of the wall-corrected RPY tensor \cite{swan_simulation_2007}.

The action of $\t{M}^{1/2}$ on the random vector $\v{W}$ is obtained through the Lanczos algorithm \cite{chow_preconditioned_2014}. To achieve convergence, the Krylov subspace $\v{K}$ is enriched iteratively with basis vectors that are linear combinations of the powers of the mobility matrix times the random vector: $\v{K}=\mbox{span}\left\{\v{W},\t{M}\v{W},\t{M}^2\v{W},...\right\}$.
The cost of the method therefore depends on the number of basis vectors, and thus mobility-vector products, required to reach a given tolerance $\epsilon$.   

\subsection{Fluctuating Force Coupling Method}
\label{sec:fluct_FCM}
The other approach relies on a matrix-free method called fluctuating Force Coupling Method (FCM) \cite{keaveny_fluctuating_2014}, which combines FCM \cite{maxey_localized_2001} with fluctuating hydrodynamics \cite{Landau1959}.
With fluctuating FCM, the coupling between the discrete degrees of freedom and the fluid is achieved through a forcing term added to the fluctuating Stokes equations,
\begin{align}
 \vnabla p - \eta \vnabla^2\v{u} &= \vnabla\cdot\t{W} + \v{f}, \label{eq:fluc_stokes1}\\
 \vnabla\cdot\v{u} &= 0,
 \label{eq:fluc_stokes2}
\end{align}
where $\eta$ is the dynamic fluid viscosity, $\v{u}$ the fluid velocity and $p$ is the pressure.

The first term in the RHS of \cref{eq:fluc_stokes1} is the divergence of the fluctuating stress tensor, $\t{W}$. $\t{W}$ is delta correlated in time and space with the following statistics,
\begin{align}
 \left\langle W_{\alpha \beta}(\v{x},t)\right\rangle &= 0,\\
  \left\langle W_{\alpha \beta}(\v{x},t)W_{\gamma \chi}(\v{y},t')\right\rangle &= 2\eta k_BT\left(\delta_{\alpha \gamma}\delta_{\beta \chi} + \delta_{\alpha \chi}\delta_{\beta \gamma} \right)\delta(\v{x}-\v{y})\delta(t-t').
\end{align}

The second term on the RHS of \cref{eq:fluc_stokes1} is the forcing transferred to the fluid with a spreading operator $\t{S}$,
\begin{align}
 \v{f}(\v{x}) = \t{S}[\v{\lambda}](\v{x}) = \sum_{i=1}^{N_b}\v{\lambda}_i\Delta_i\left(\v{x}\right),
 \label{eq:spreading}
\end{align}
where the finite size of the discrete degrees of freedom is accounted for with a Gaussian spreading envelope,
\begin{align}
 \Delta_i(\v{x}) = (2\pi\sigma^2)^{-3/2}\exp\left(-\frac{\|\v{x}-\v{r}_i\|^2}{2\sigma^2}\right).
\end{align}
The length scale $\sigma$ functions in a similar way to the bead radius, $a$, in the RPY tensor.  The diagonal entries of the FCM mobility matrices will match those of the RPY tensor if $\sigma = a/\sqrt{\pi}$ \cite{maxey_localized_2001}.

The fluid velocity, obtained after solving \cref{eq:fluc_stokes1,eq:fluc_stokes2}, is the sum of a deterministic part, due to the forcing $\v{\lambda}$, and a fluctuating term, stemming from the fluctuating stress.  We may express the total fluid velocity as
\begin{align}
 \v{u} &= \t{L}^{-1}\left(\t{S}\v{\lambda} + \t{D}\t{W}\right)\\
 &= \v{u}^D + \breve{\v{u}},
\end{align}
 where $\t{S}$ is the spreading operator in \cref{eq:spreading}, $\t{L}^{-1}$ is the inverse Stokes operator (i.e.\ the fluid solver), and $\t{D}$ the divergence operator applied to the fluctuating stress \cite{Delong2014}.

The velocities of the discrete degrees of freedom are then obtained from the fluid velocity using an averaging operator, $\t{J}$, such that
\begin{align}
 \v{v}_i = (\t{J}[\v{u}])_i = \int \v{u}\Delta_i(\v{x}) \d^3\v{x} = \v{v}_i^D + \breve{\v{v}}_i, \quad i=1,..,N_b,
 \label{eq:averaging}
\end{align}
where $\t{J} = \t{S}^\star$ is adjoint to the spreading operator.

We see then that the FCM mobility matrix can be written as the composition of three linear  operators: $\t{M}_{FCM} = \t{J}\t{L}^{-1}\t{S}$.  Additionally, as demonstrated in \cite{keaveny_fluctuating_2014-1} the velocity $\breve{\v{v}}$ satisfies the fluctuation–dissipation theorem with the covariance given by the FCM approximation of the mobility matrix, 
\begin{align}
 \left\langle \breve{\v{v}}(t)\breve{\v{v}}(t') \right\rangle= 2k_BT\t{M}_{FCM}\delta(t-t'),
 \label{eq:fluctu_dissip}
\end{align}
where $\breve{\v{v}} = \sqrt{2k_BT}\t{M}^{1/2}_{FCM}\v{W}(t)$, with $\t{M}^{1/2}_{FCM} =\t{J}\t{L}^{-1}\t{D}$.

Thus, we see that fluctuating FCM simultaneously computes the actions of $\t{M}$ and $\t{M}^{1/2}$ on $\v{\lambda}$ and $\v{W}$, respectively, by considering solutions to the forced fluctuating Stokes equations.

Following \cite{keaveny_fluctuating_2014-1,delmotte_simulating_2015}, we implement fluctuating FCM in periodic domains.  This involves first evaluating the forcing, $\v{f}(\v{x})$, and the fluctuating stress, $\t{W}$, on a regular grid.  Next, the Stokes equations are solved using a Fourier spectral method.  Finally, the trapezoidal rule is used to numerically integrate \cref{eq:averaging} to obtain the translational velocity for each of the discrete degrees of freedom making up the rigid bodies.

\section{Computational cost of the gDC}
\label{sec:gDC_cost}

Due to the differences in their implementations, there is a distinct difference in cost between RPY and grid-based fluctuating FCM that manifests itself when computing the random velocity vector, $\breve{\v{v}}$.  For both implementations, the most expensive computation is applying the matrix $\t{M}$.  With RPY, as we are directly handling $\t{M}$, any time the particle positions change, we must compute $\breve{\v{v}} = \t{M}^{1/2}\v{W}$ using the Lanczos algorithm, for which each iteration requires a matrix-vector multiplication involving $\t{M}$.    

For fluctuating FCM, however, we are instead working with a decomposition of $\t{M}_{FCM}^{1/2}$ such that $\breve{\v{v}} = \sqrt{2k_BT}\t{J}\t{L}^{-1}\t{D}\t{W}(t)$.  In this decomposition, the only matrix that depends on the particle configuration is $\t{J}$.  Thus, provided that $\t{W}$ remains the same, changing the particle positions simply requires reaveraging the fluid velocity $\breve{\v{u}}$ at the new positions, and avoids having to recompute the velocity field itself.  






\noindent \textit{Cost of Step 2:} Along with the computation of $\breve{\v{v}}$, Step 2 of the gDC algorithm in \cref{sec:gDC_cost} involves solving a simplified mobility problem with $\t{M}$ replaced by $\t{I}$.  This is equivalent to solving the least squares problem $\min_{\v{\mathcal{V}}}\left\|\t{K}\v{\mathcal{V}} - \breve{\v{v}}\right\|$, the solution of which is given by \cref{eq:sol_unconst} and involves the $6N\times6N$ symmetric, block-diagonal matrix $\t{K}^\top \t{K}$.   The $6\times 6$ block associated with body $p$ is given by
\begin{align}
 (\t{K}^\top \t{K})^{(p)} = \left[\begin{array}{cc}
N_b^{(p)}\t{I}_3 & \sum\limits_{i\in\mathcal{B}_p}\left[\times (\v{r}_i - \v{Y}_p)\right]\\ 
\sum\limits_{i\in\mathcal{B}_p}\left[(\v{r}_i - \v{Y}_p)\times\right] &  -\sum\limits_{i\in\mathcal{B}_p}\left[(\v{r}_i - \v{Y}_p)\times\right]^2\\
\end{array}\right],
\end{align}
where $\mathcal{B}_p$ is the set of discrete degrees of freedom belonging to body $p$.
Therefore, computing $\left(\t{K}^\top \t{K}\right)^{-1}$ just requires inverting $N$ small  $6\times6$ blocks using dense linear algebra, which is negligible in terms of computational cost and, additionally, naturally lends itself to parallel computation.

Thus, the primary cost associated with Step 2 involves computing $\breve{\v{v}}$, which for fluctuating FCM incurs the cost of one Stokes solve and $M$ averaging operations.  The RPY implementation will require $M_{\textrm{Lanczos}}$ mobility-vector products with the Lanczos method. 

\noindent \textit{Cost of Step 3:} The main cost of this step is computing the scalar term $\nu$ using finite differences (FD) or RFD to generate $\partial_\v{x}\cdot \v{\mathcal{V}}$.  Finite differencing  involves perturbing the positions and orientations of each rigid body and calculating the new fluctuating velocities for the discrete degrees of freedom at the disturbed states. With RFD, the new fluctuating velocities are only calculated once, after disturbing the particles' positions and orientations with a random vector $\widetilde{\v{W}}$.

For fluctuating FCM, FD is straight forward and incurs minimal cost.  Since the body velocities are uncorrelated in \cref{eq:unconst}, the finite-differencing step in \cref{eq:FD} requires $6M$ averaging operations.  As discussed above, however, changing the particle position with the RPY-based implementation will require recomputing $\t{M}^{1/2}\v{W}$ with each displacement.  Thus, we require $6N\times M_{\textrm{Lanczos}}$ mobility-vector products.  This is a cost that could rise quickly as the particle number increases. 
On the other hand, using RFD in \cref{eqn:rfd_div}  only requires $M_{\textrm{Lanczos}}$  mobility-vector products  with matrix-based methods and $M$ averaging operations with fluctuating FCM.

\noindent \textit{Cost of Step 5:} This step involves the costliest computation of solving the full mobility problem, \cref{eq:lin_sys_RMB}, along with the additional computation of $\breve{\v{v}}$.  

For the fluctuating FCM implementation, each GMRES iteration requires a full fluctuating FCM computation that involves spreading the force vector $\v{\lambda}$, solving the Stokes equations, and finally averaging the resulting flow field to obtain the velocities for the discrete degrees of freedom.  A single additional averaging is needed to incorporate the fluctuating velocities, $\breve{\v{v}}^m$, on the RHS of \cref{eq:step5}.  

For matrix-based methods, each GMRES iteration will incur the cost of a matrix-vector product involving $\t{M}$.  There will also be the cost of $M_{\textrm{Lanczos}}$ Lanczos interations needed to obtain $\breve{\v{v}}^m$.  

For a tolerance of $\epsilon = 10^{-3}$, both the Lanczos and GMRES with block preconditioning require approximately $M_{\textrm{Lanczos}} = M_{\textrm{GMRES}} = 5$ interations on average.  The tolerance of $\epsilon = 10^{-3}$ has been shown \cite{sprinkle_large_2017} to be a suitable choice for simulations involving fluctuations.  Adding these costs together assuming $\epsilon = 10^{-3}$, we see that the cost for one gDC-FD time iteration with fluctuating FCM is ($1+5=$) 6 Stokes solves, $5M$ spreading operations and ($1+6+1+5=$) $13M$ averaging operations.  For the RPY based implementation, we have ($5 + 6N\times5 + 5 + 5=$) $5\times(6N+3)$ matrix-vector products involving $\t{M}$. With gDC-RFD the number of averaging operations reduces to $8M$ for fluctuating FCM, and the total number of matrix-vector products involving $\t{M}$ with the RPY based implementation lowers to $20$,  becoming independent of the particle number $N$. 

As shown in \cref{sec:convergence}, the gDC is more stable  with FD than with RFD. For grid-based methods, the $5M$ additional averaging operations incurred by FD are negligible compared to the Stokes solve. This is not true for matrix based methods for which RFD represents a significant cost reduction.

Based on these costs, we see that the gDC with FD is well-suited for grid-based methods utilising fluctuating hydrodynamics such as fluctuating FCM \cite{keaveny_fluctuating_2014,delmotte_general_2015}, PSE \cite{Fiore2017,Fiore2018} or the fluctuating and stochastic Immersed Boundaries methods \cite{Atzberger2007,Atzberger2011,Delong2014,Sprinkle2019}, while  gDC with RFD is the best compromise between numerical stability and computational cost for matrix-based methods.


\section{Simulations}
In this section, we demonstrate the performance of the gDC by performing simulations of particulate suspensions under both dynamic and equilibrium conditions.  In  \Cref{sec:convergence}, we investigate the accuracy of the gDC by examining the equilibrium distributions of the position and orientation of a boomerang-shaped particle in a gravitational field above a planar no-slip boundary.  Owing to the small size of this system, we can integrate for long times and acquire many realisations to quantify the temporal accuracy of the gDC.  We compare these results with those obtained using the state-of-the-art integrators developed in Sprinkle \textit{et al.} \cite{sprinkle_large_2017}.  After this, we demonstrate the suitability of the gDC for large scale simulations of anisotropic Brownian particles. In  \Cref{sec:crystal}, we study the equilibrium properties of confined suspensions of rod-shaped particles, a model for a liquid crystal or stiff-polymer system.  Finally, inspired by recent experimental work \cite{Bourrianne2020}, in  \Cref{sec:rheology} we use the gDC to perform simulations to obtain the rheology of colloidal suspensions made from Czech hedgehog-shaped (CH) particles. 

\subsection{Implementations}
\label{sec:impl}
The simulations that follow rely on two implementations: one using direct evaluations of the wall-corrected RPY tensor and the other using fluctuating FCM.
\begin{itemize}
    \item \textbf{Python with RPY:} for the single boomerang simulations we use the collaborative code, called ``RigidMultiblobWall", that one of us co-developed \cite{usabiaga_hydrodynamics_2016} and used for large scale simulations of colloidal active particles \cite{driscoll_unstable_2017,Balboa2017}. The code contains most of the time integration schemes recently developed for matrix-based approaches \cite{delong_brownian_2015, Balboa2017,sprinkle_large_2017}. It relies on the pairwise evaluation of the wall-corrected RPY tensor. The mobility computations can be accelerated in different ways: using C++ routines,  Numba or GPUs via the PyCUDA interface.  We implemented the gDC in the code to compare its performance with the most accurate scheme developed by Sprinkle \textit{et al.}: the Trapezoidal Slip scheme (T-S).  
    \item \textbf{C++ with fluctuating FCM:} for the confined liquid crystal and rheology simulations, we use an FFT-based fluid solver with fluctuating FCM in a periodic box. FFTs are parallelised with the scalable MPI library FFTW. MPI is also used to parallelise the spreading and averaging operations as well as the nearest neighbour search necessary to compute short-ranged, pairwise interactions between discrete degrees of freedom. At each grid point, the entries of the fluctuating stress are independent Gaussian random variables. Since the fluctuating stress $\t{W}$ is symmetric, at each grid point we generate six random numbers based on a Gaussian distribution with zero mean and unit variance. We then multiply the off-diagonal entries by $\sqrt{2k_BT\correction{\eta}/(\Delta x^3)}\Delta t^{-1/2}$ and the diagonal ones by $2\sqrt{k_BT\correction{\eta}/(\Delta x^3)}\Delta t^{-1/2}$. We truncate the Gaussian envelopes by setting $\Delta_i(\v{x})=0$ for $\|\v{x}-\v{r}_i\|> 3a$, and the length scale of the envelope, $\sigma$, is related to the grid size, $\Delta x$, through $\sigma/\Delta x = 1.86$. For more details, we refer the reader to our previous work \cite{delmotte_simulating_2015}.
\end{itemize}


\subsection{Convergence study: boomerang above no-slip boundary}
\label{sec:convergence}

\begin{figure}[htb!]
    \centering
    \includegraphics[width=0.45\textwidth]{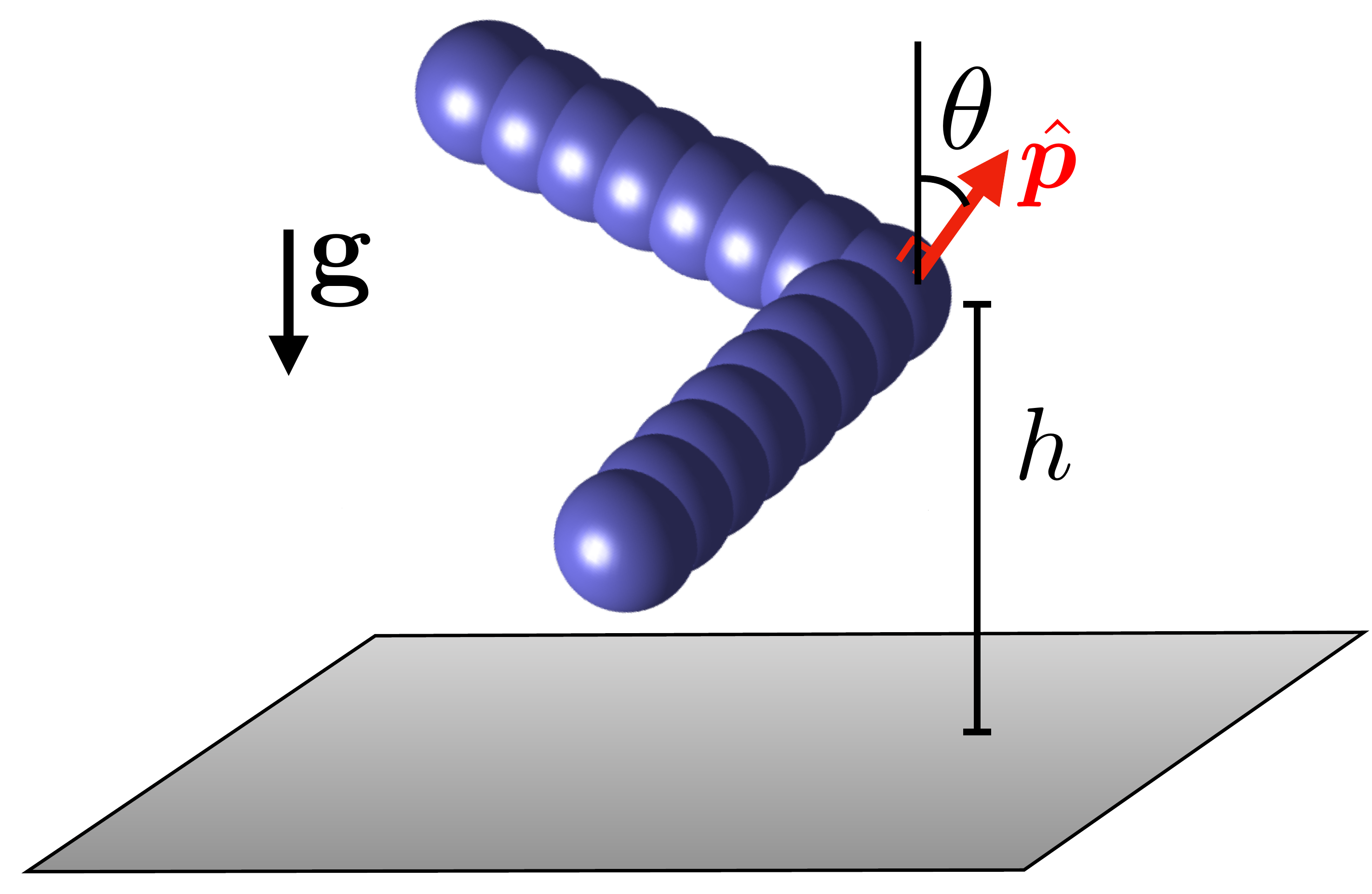}
    \caption{Colloidal boomerang made of 15 RPY-particles above a no-slip boundary. The height, $h$, is the distance between the centre of the boomerang-point RPY-particle and the no-slip surface.  The angle $\theta$ is that between the vertical axis and the unit vector, $\hat{\v{p}}$, that is orthogonal to both boomerang arms.}
    \label{fig:boom_schematic} 
\end{figure}

In this section we study the accuracy of the gDC scheme by examining the equilibrium distributions of a boomerang particle in a gravitational field above a no-slip boundary. Micron-size boomerangs have been used extensively to study the diffusion of colloidal anisotropic particles in experiments \cite{Chakrabarty2013} and simulations \cite{delong_brownian_2015,usabiaga_hydrodynamics_2016}.  Following Delong \textit{et al.} \cite{delong_brownian_2015} and as shown in  \cref{fig:boom_schematic}, we discretise the boomerang using 15 RPY-particles separated by distance $a$, the radius of the RPY-particle.  The boomerang is subject to a gravitational force of magnitude $mg = 0.18k_BT/a$ and each RPY-particle interacts with the planar boundary through a short-ranged, repulsive potential, 
\begin{equation}
        V(h) =  V_0 a \frac{\exp\left(-\frac{h - a}{b}\right)}{h-a},
\end{equation}
where $h$ is the height of the RPY-particle's centre from the surface,  $V_0 = 23k_BT$ sets the potential strength, and $b = 0.5a$ is the range of the potential.

Simulations are performed using both the T-S and gDC  schemes for three different time-step sizes $\Delta t = 0.1, 0.2, 0.3\tau_D$, where $\tau_D = a^2/D = 6\pi\eta a^3/k_BT$ is the typical diffusive timescale associated with an RPY-particle. The divergence term, $\nu$, in Step 3 of the gDC algorithm (see \cref{sec:gDC_alg}) is computed using both FD \eqref{eq:FD} and RFD  \eqref{eqn:rfd_div}.  The solver tolerance for the GMRES and Lanczos algorithm is set to $10^{-4}$,
 and the finite difference parameter is $\delta = 10^{-6}$.

In order to obtain sufficient statistics, we run 36 different simulations for each value of $\Delta t$.  Each simulation is initialized using a Monte Carlo generated sample of the Gibbs-Boltzmann distribution and is run to a final time of $30000\tau_D$, where the solution is recorded at every $t = 0.3\tau_D$.

\begin{figure}[htb!]
    \centering
    \includegraphics[width=0.45\textwidth]{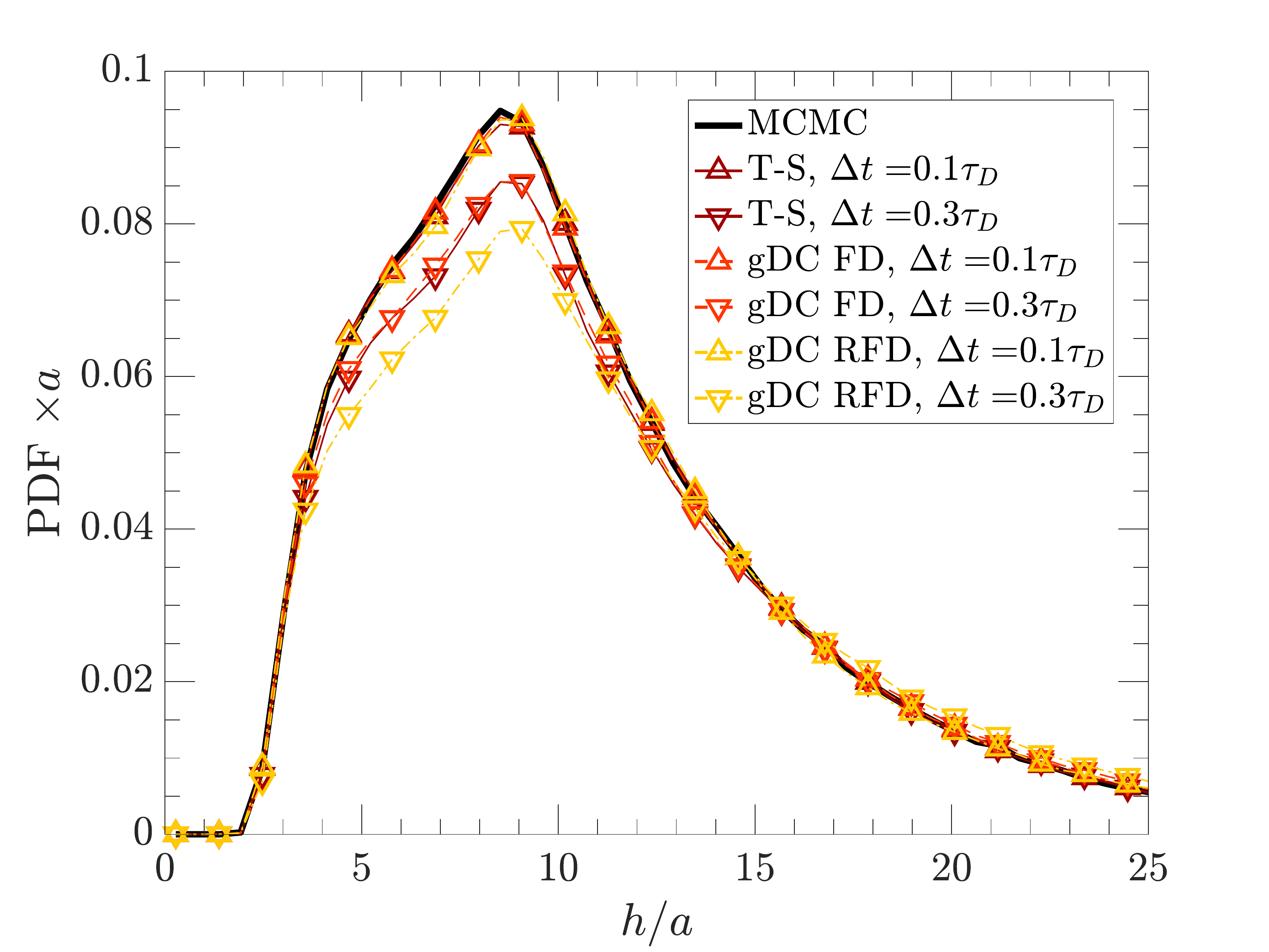}
    \includegraphics[width=0.45\textwidth]{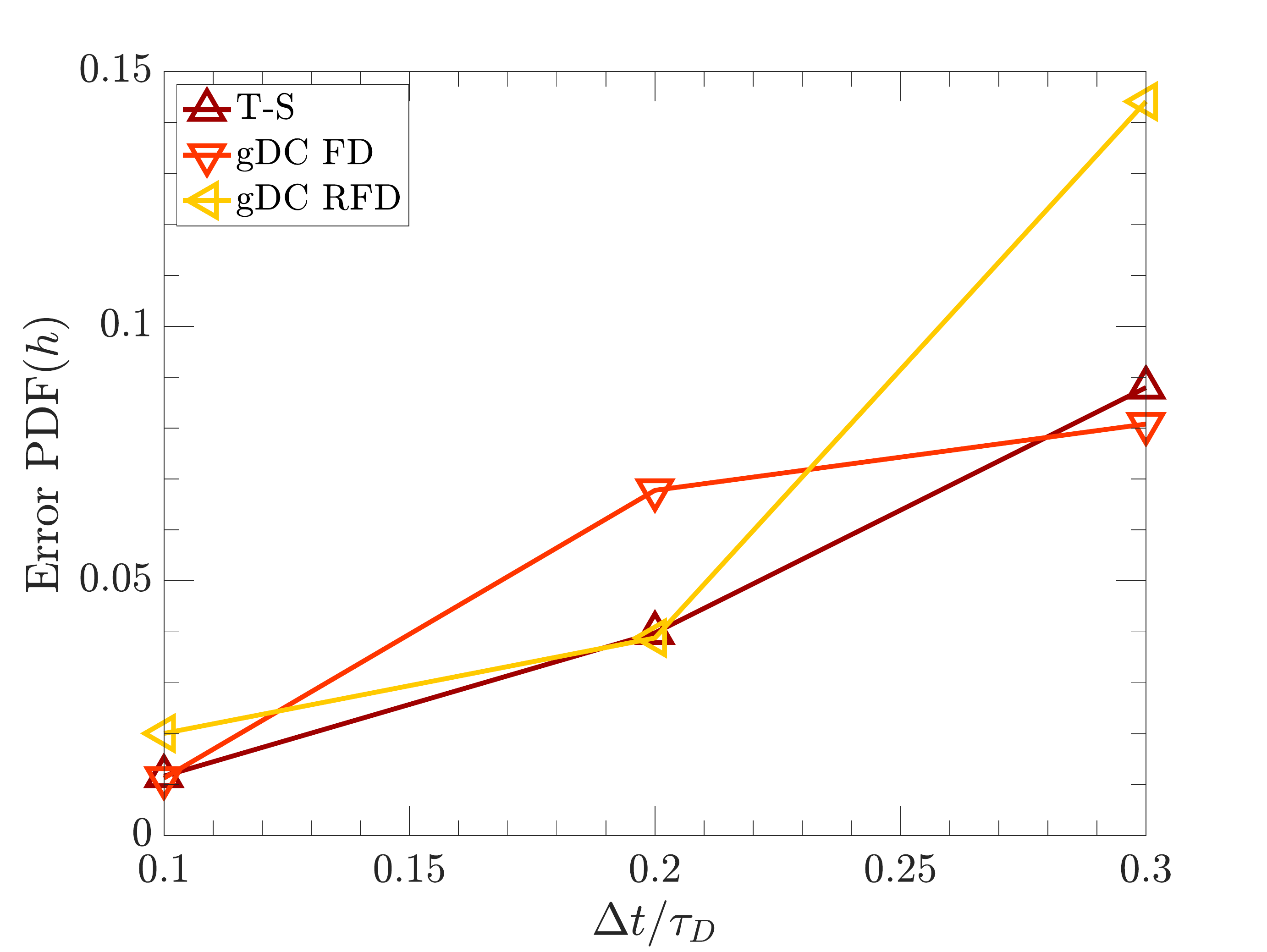}
    \caption{(left panel) Height distribution of the boomerang above a no-slip surface (right panel) $L_2$-error between the simulated distributions and the true distribution generated using MCMC. }
    \label{fig:single_boom_distributions} 
\end{figure}

\Cref{fig:single_boom_distributions} compares the height distributions from the T-S and gDC simulations with the true distribution obtained using Markov Chain Monte Carlo (MCMC). We see that, as expected, both schemes converge to the MCMC distribution as $\Delta t$ decreases. The $L_2$ errors for the distributions shown in the right panel indicate that the gDC achieves a similar accuracy and convergence rate as the T-S. However,  the RFD computation of the $\nu$ term in the gDC algorithm leads to larger errors than FD. This is particularly true for the largest time-step, $\Delta t = 0.3 \tau_D$, for which gDC-RFD exhibit numerical stability issues due to particle overlaps across the wall. Similar stability problems were also encountered with the T-S scheme, resulting in several simulations being discarded and restarted when $\Delta t = 0.3 \tau_D$.  We suspect that the loss of stability with gDC-RFD is due to  overestimations of $(\partial_\v{x}\cdot \v{\mathcal{V}})^n$ caused by the random sampling $\widetilde{\v{W}}$, which leads to large values of the corrective term $\nu$, therefore potentially moving the particle too far into the wall at the next time step. 

\begin{figure}[htb!]
    \centering
    \includegraphics[width=0.45\textwidth]{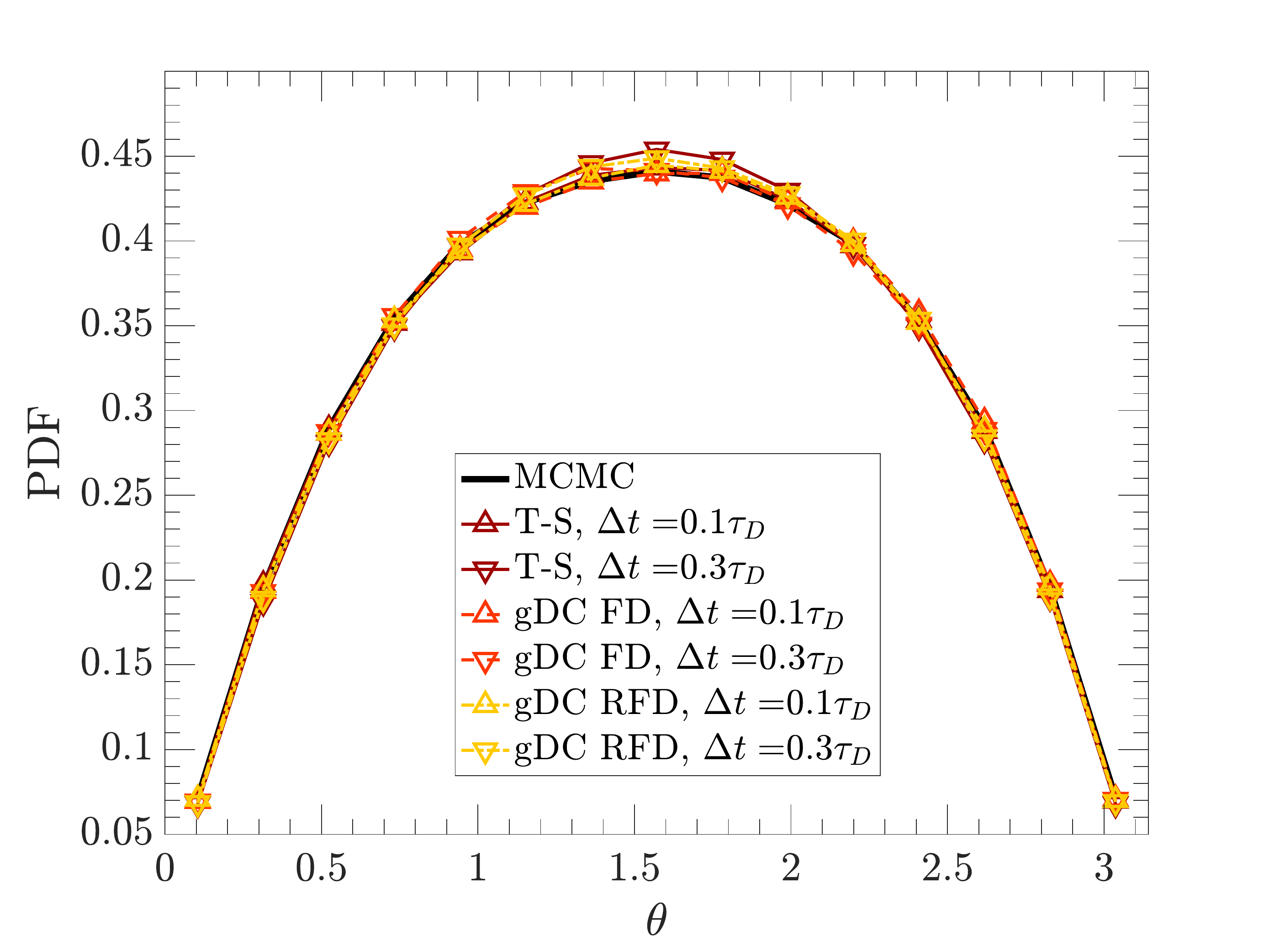}
    \includegraphics[width=0.45\textwidth]{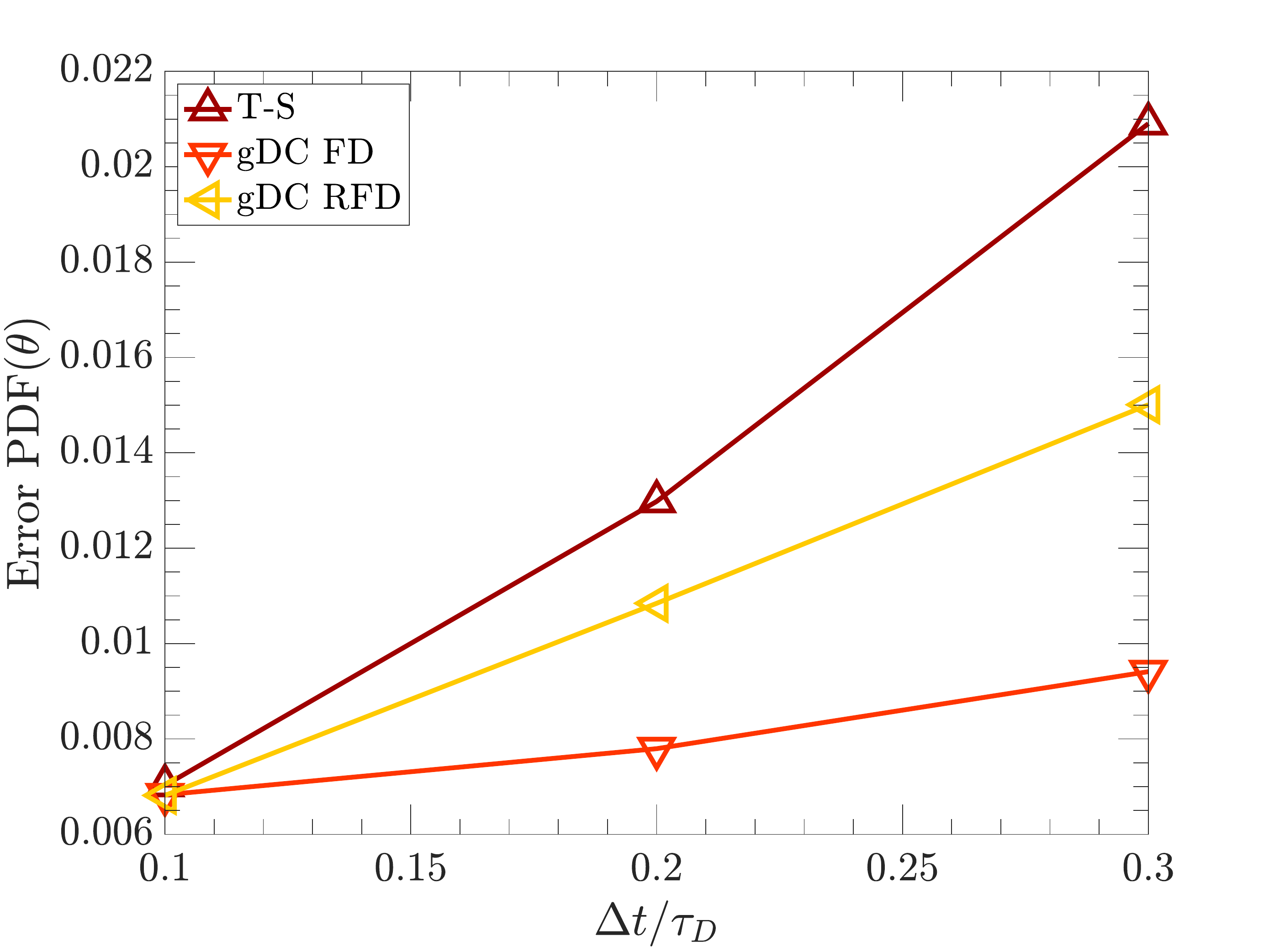}
    \caption{(left panel) Orientational distribution of the boomerang above a no-slip surface (right panel) $L_2$-error between the simulated distributions and the true distribution generated using MCMC. }
    \label{fig:single_boom_distributions_2} 
\end{figure}

The orientation distribution, shown in  \cref{fig:single_boom_distributions_2}, displays a similar trend, though the gDC exhibits a smaller error which does not change considerably with the time-step size.  

\subsection{Confined liquid crystal}
\label{sec:crystal}

\begin{figure}[htb!]
    \centering
    \includegraphics[width=0.49\textwidth]{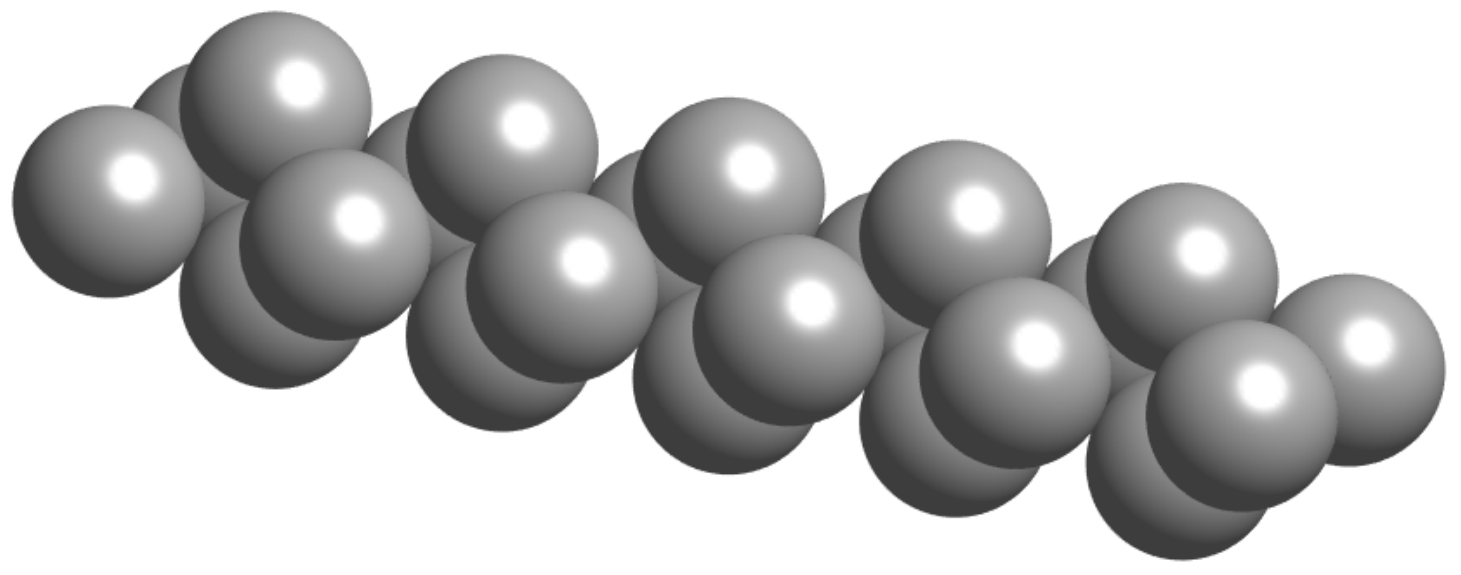}
    \caption{A rod formed of 22 FCM-particles.}
    \label{fig:rod_shaped_body} 
\end{figure}

Inspired by the alignment of slender molecules in liquid crystals, this section examines the equilibrium distributions of confined suspensions of rod-shaped particles. In this set of simulations, each rod is constructed from 22 particles of radius $a$ as illustrated in \cref{fig:rod_shaped_body}. The simulations are performed using fluctuating FCM in periodic computational domains with dimensions $[0,L_x]\times[0,2L_y]\times[0,L_z]$. Confinement is introduced by first including slip boundaries at $y = 0$ and $y = L_y$ through boundary conditions $\v{u}\cdot\uv{e}_y = 0$ and $\left(\t{I} - \uv{e}_y\uv{e}_y^\top\right)\vdel\v{u} = \vu{0}$ on the flow field.  These conditions are enforced using an image system \cite{delmotte_simulating_2015} for both the forces on the bodies, as well as the fluctuating stress, in the $y > L_y$ half of the computational domain. Second, the rods are kept away from the channel walls by including a repulsive harmonic potential
\begin{equation}
    V(r) = \left\{\begin{matrix}\frac{K}{2}\left(r-b\right)^2, & r < b, \\ 0, & r \geq b,\end{matrix}\right.
\end{equation}
on each FCM-particle making up the rod.  Here, $r$ is the distance between an FCM-particle centre and the wall, $b = 2.2a$ is the cut-off distance, and $K = \frac{40k_BT}{ab}$.  Along with this potential, overlap between FCM-particles is also discouraged using the soft potential
\begin{equation} \label{eqn:soft_sphere_potential}
        U(r) = \left\{\begin{matrix}U_0\left(1 + \frac{2a - r}{b}\right), & r < 2a, \\ U_0\exp\left(\frac{2a - r}{b}\right), & r \geq 2a,\end{matrix}\right.
\end{equation}
with $r$ now denoting the centre-to-centre distance between 2 FCM-particles, $U_0 = 20k_BT$ and $b = 0.5a$.

\begin{figure}[htb!]
    \centering
    \includegraphics[width=0.5\textwidth]{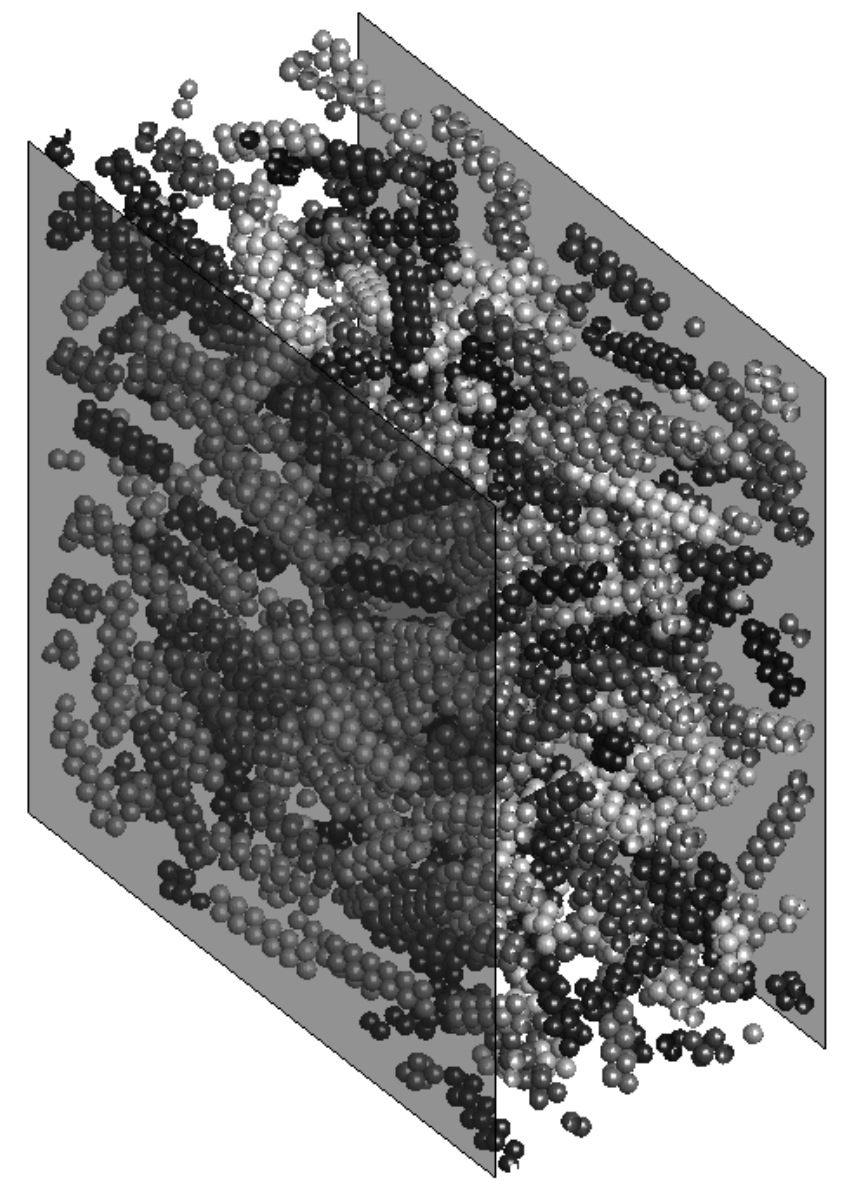}
    \caption{A snapshot from a rod simulation with channel width $L_y = L/2$. Each rod has a randomly-selected greyscale colouring and the semi-transparent grey boundaries depict the locations of the slip boundaries.}
    \label{fig:rod_sim_snapshot} 
\end{figure}

In the following simulations, the domain length is set to $L_x = L_z = L = 77.65a$. We consider three different channel widths, namely $L_y = L/8$, $L_y = L/4$ and $L_y = L/2$, to explore how confinement impacts the rod distributions. The number of rods in these simulations is taken to be 64, 128 and 256 respectively to maintain a constant volume fraction of 10\%. A representative simulation snapshot shown in \cref{fig:rod_sim_snapshot}. Each simulation was run until a final time of $50\tau_D/3$ and the time step size was set to $\Delta t = \tau_D/600$. Finally, the initial conditions for the gDC simulations were generated using samples from independent MCMC runs.

\begin{figure}[!p]
    \centering
    \includegraphics[width=0.49\textwidth]{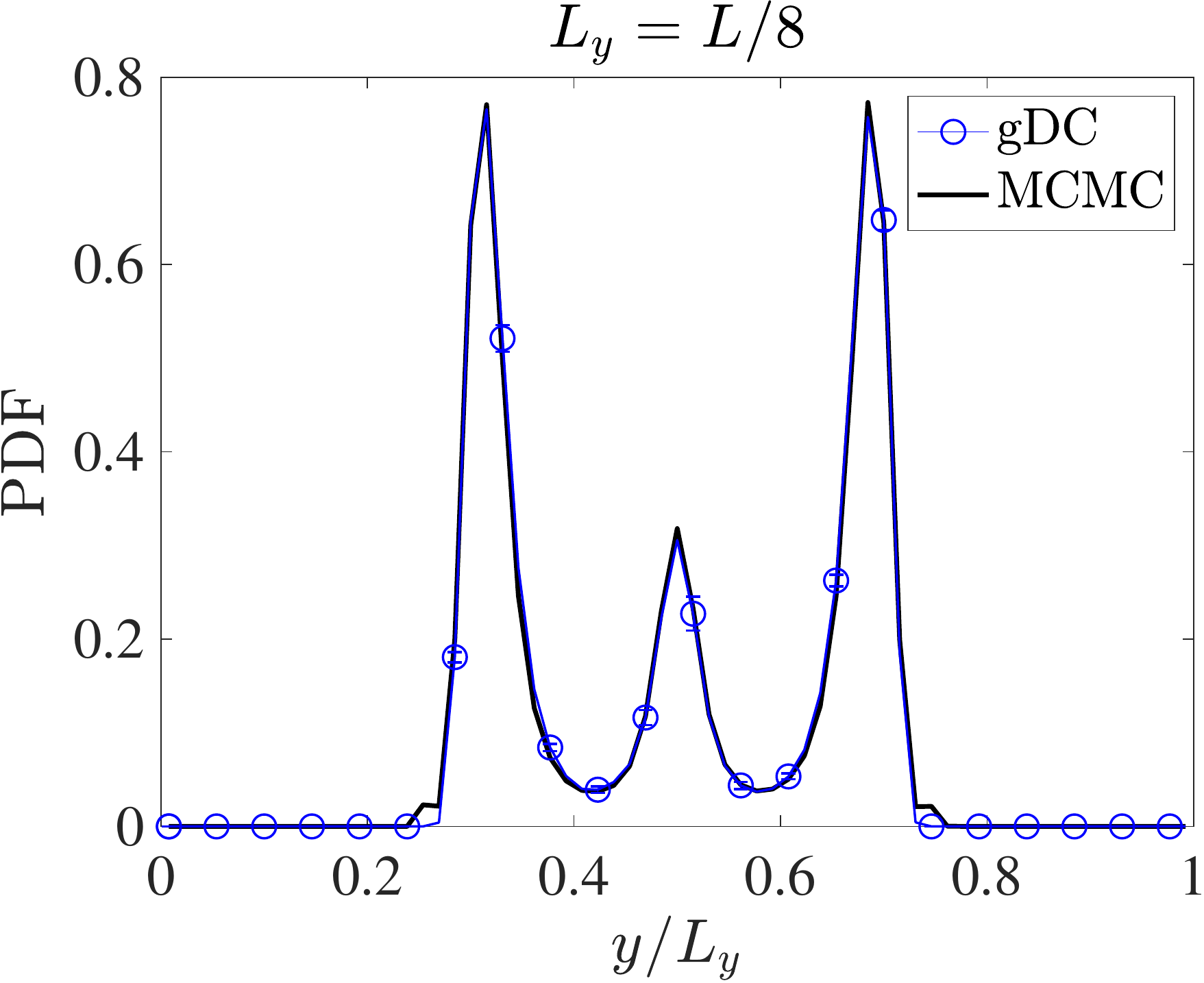}
    \includegraphics[width=0.49\textwidth]{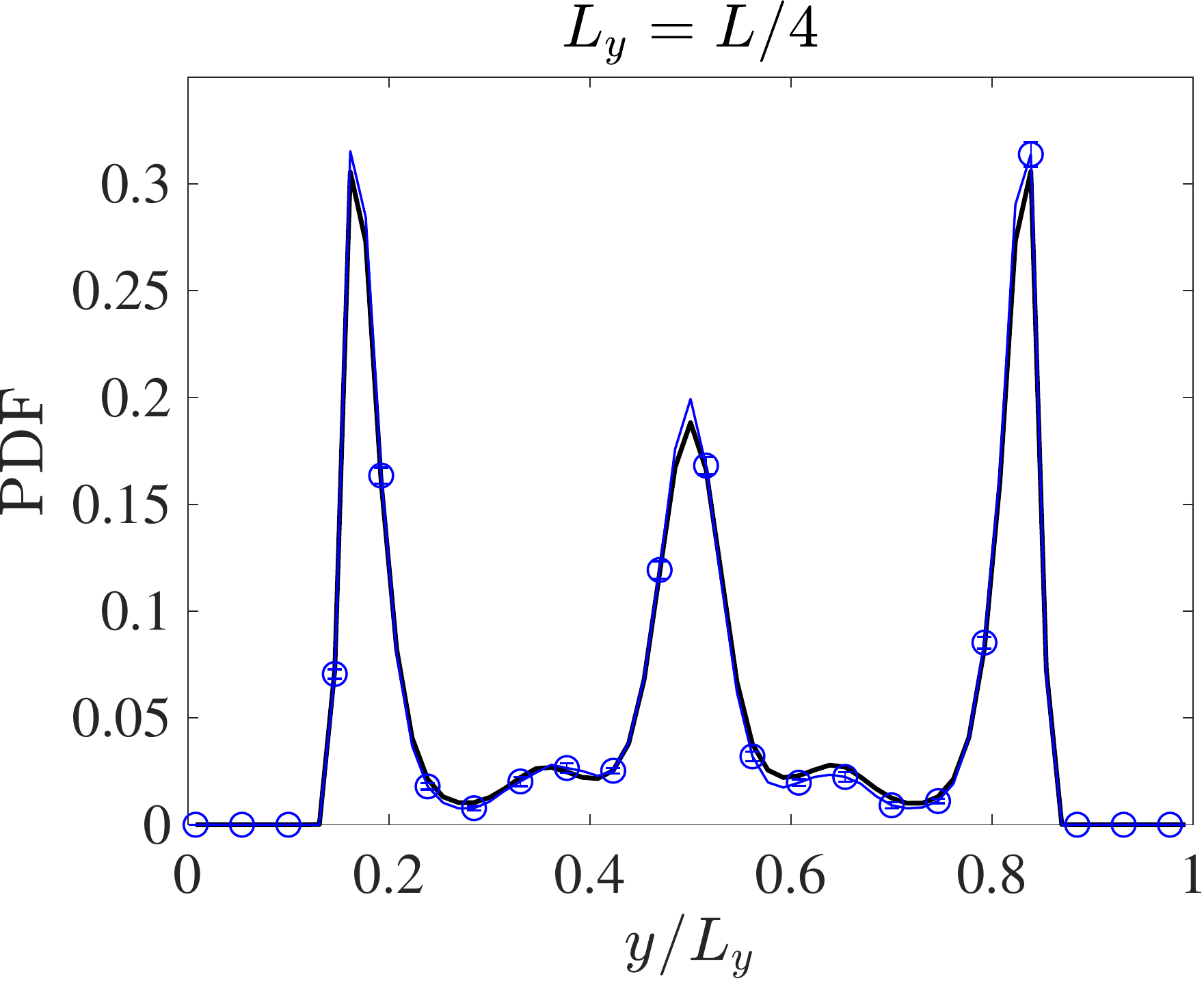}
    \includegraphics[width=0.49\textwidth]{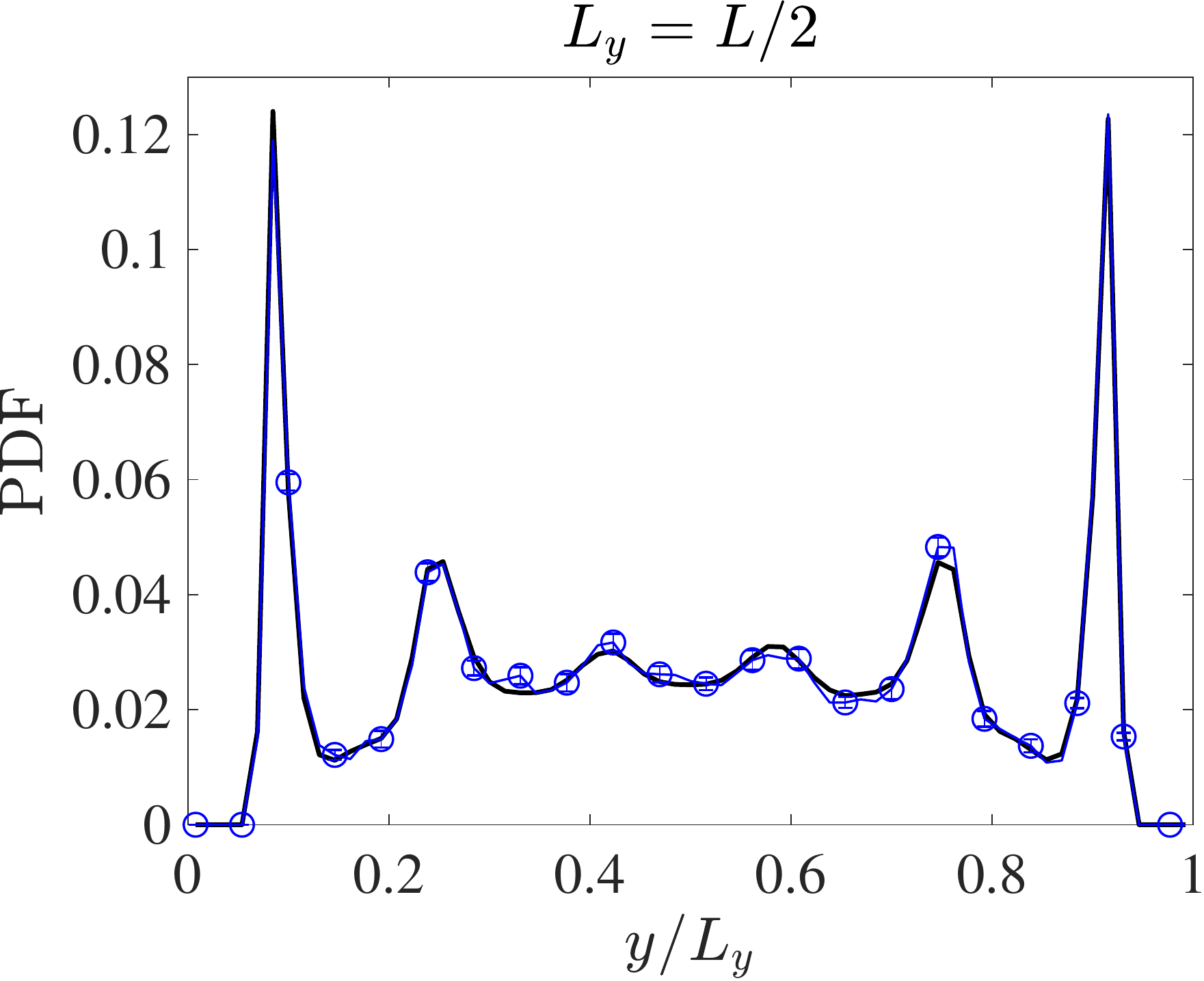}
    \caption{The equilibrium distributions of rod positions across slip channels of widths $L_y = L/8$, $L_y = L/4$ and $L_y = L/2$.}
    \label{fig:channel_y_distributions}
\end{figure}

\begin{figure}[!p]
    \centering
    \includegraphics[width=0.49\textwidth]{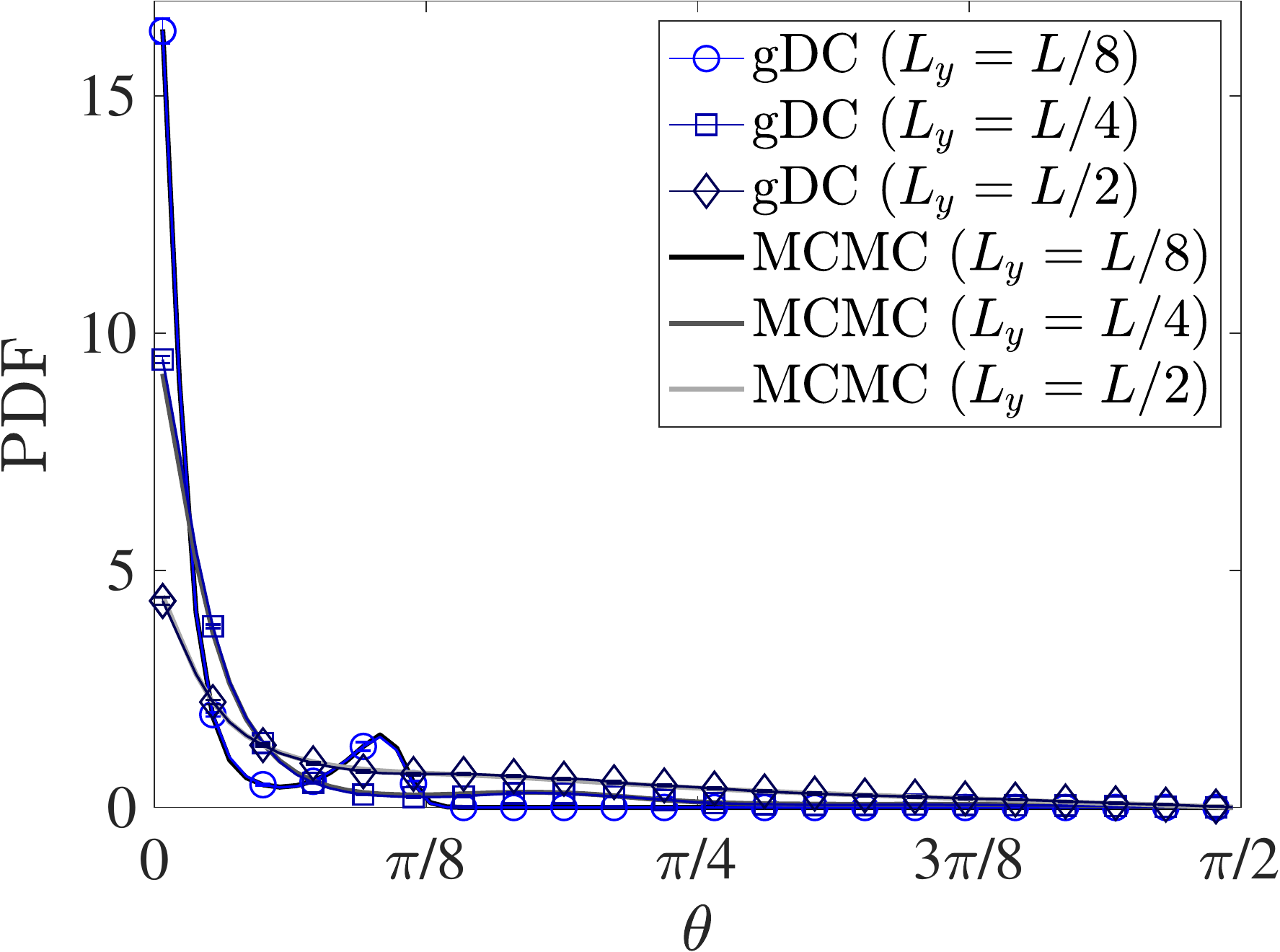}
    \includegraphics[width=0.49\textwidth]{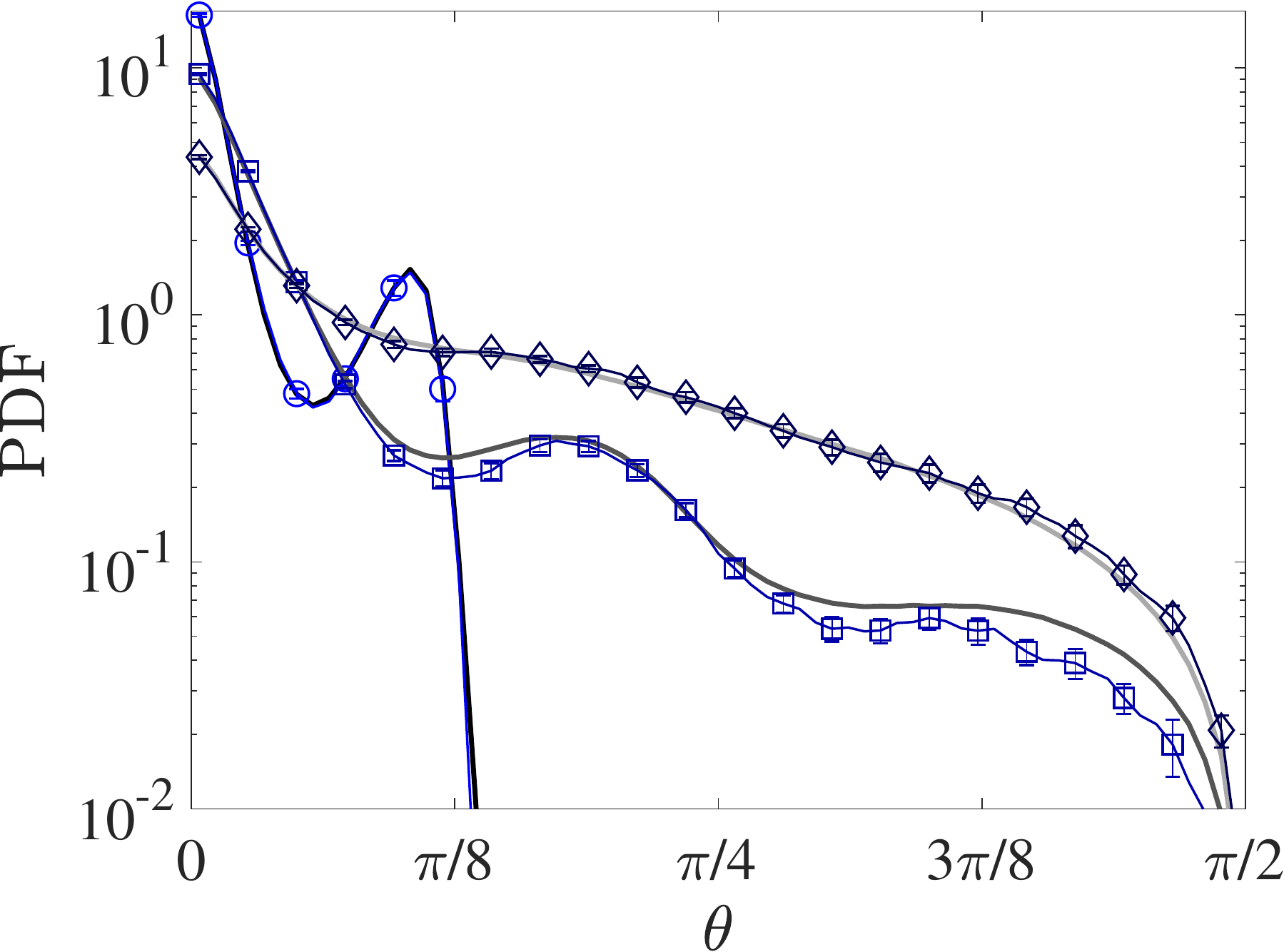}
    \caption{The equilibrium distributions of rod orientations in slip channels of various widths, shown in both (left) linear and (right) logarithmic scales.}
    \label{fig:rod_angle_distributions}
\end{figure}

To measure the distribution of the rods within the channel, we compute the distribution of the $y$-coordinates of geometric centre of the rods.  Additionally, we compute the distribution of the angle $\theta$ between the rod-axis and the $xz$-plane, such that $\theta = 0$ corresponds to the rod being parallel to the channel boundaries.  \Cref{fig:channel_y_distributions} compares the $y$-coordinate distributions from the gDC simulations with the Gibbs-Boltzmann distributions calculated using MCMC. For all channel widths that were considered, good agreement is observed between these distributions. Similarly close agreement is found for the $\theta$-distributions in \cref{fig:rod_angle_distributions}; the shape of the distribution is well-preserved even when viewed on a logarithmic scale. All of the $y$-coordinate distributions exhibit the expected symmetry about the midpoint of the channel at $y = L_y/2$, and have peaks in probability density near the slip boundaries. The position of the peaks depends on the width of the slip channel because of the finite thickness of the rods. It is also clear that the narrower the channel, the higher the probability of the $y$-positions associated with these peaks. The $\theta$-distributions all decrease from a maximum value at $\theta \approx 0$, although for the narrowest channel, the distribution exhibits a second local maximum just below $\theta = \pi/8$. The wider the channel, the more slowly the probability density decays, and hence the larger the range of likely orientations.

For the narrowest channel, the channel width, $L_y = L/8 \approx 9.7a$, is slightly larger than twice the diameter of the rod cross-section ($2 \times 4a = 8a$), but less than the rod length of $16.14a$.  As a result, the harmonic potential which keeps rods away from the slip surfaces will also induce a torque on the rods even for small deviations from $\theta = 0$. Thus, the rods are most likely to be found with an orientation $\theta \approx 0$ in one of the two layers displaced from the channel centre. The distance between these layers balances the repulsive potentials between different rods and between particles and the walls.  The smaller peak at the channel centre represents the small number of rods that get `jammed' between the two layers.  Additionally, as there is insufficient space for all rods to have $\theta \approx 0$, they also tend to span the channel diagonally with their ends trapped between the channel walls, leading to the secondary peak in the $\theta$-distribution below $\theta = \pi/8$.
 
 As the channel width increases when we have $L_y = L/4 \approx 19.4a$, the central peak in the $y$-coordinate distribution becomes more pronounced, as shown in \cref{fig:channel_y_distributions}. Additionally, we observe a wider range of possible $\theta$-value for rods near the channel centre, although we note that a small, secondary peak in the $\theta$ distribution remains as seen in the log-scale plot in \cref{fig:rod_angle_distributions}. Note that this peak occurs at a larger $\theta$ value than for the narrowest channel due to the increased channel width. Rods in the outer layers between this central layer and the slip boundaries are still most likely to have $\theta \approx 0$.  As a consequence, the rods in the middle layer are also likely to align with $\theta = 0$ with the two outer rod layers acting like the channel walls.  Therefore, we see that the angle distribution retains its maximum at low $\theta$.
 
For the largest width, the $\theta$ distribution still exhibits its peak value at $\theta \approx 0$ as the rods nearest the slip boundaries continue to align parallel to the boundaries.  We again observe peaks in the $y$-coordinate distribution close to the channel boundaries, though their magnitude is now reduced as shown in \cref{fig:channel_y_distributions}. The central peak observed in the two previous cases, however, has now spread out to form a series of peaks of decreasing magnitude in the interior of the channel. Note that this persistent pattern of density peaks which are largest near the walls and which decrease towards the channel centre is consistent with existing observations on the layering of colloidal particles near repulsive boundaries \cite{eshraghi_molecular_2018,snook_monte_1978,mittal_does_2007,deb_hard_2011,van_winkle_layering_1988}. Rods appear to spend comparable amounts of time between and in these internal layers, seemingly making layers away from the slip surfaces more transient than those near the boundaries where rods tend to spend significant amounts of time. Finally, rods in the channel interior achieve a broader distribution of $\theta$ values with no discernible secondary peak in the $\theta$ distribution.

\subsection{Rheology of suspensions of Czech hedgehog colloids}
\label{sec:rheology}

\begin{figure}[!t]
    \centering
    \includegraphics[width=0.39\textwidth]{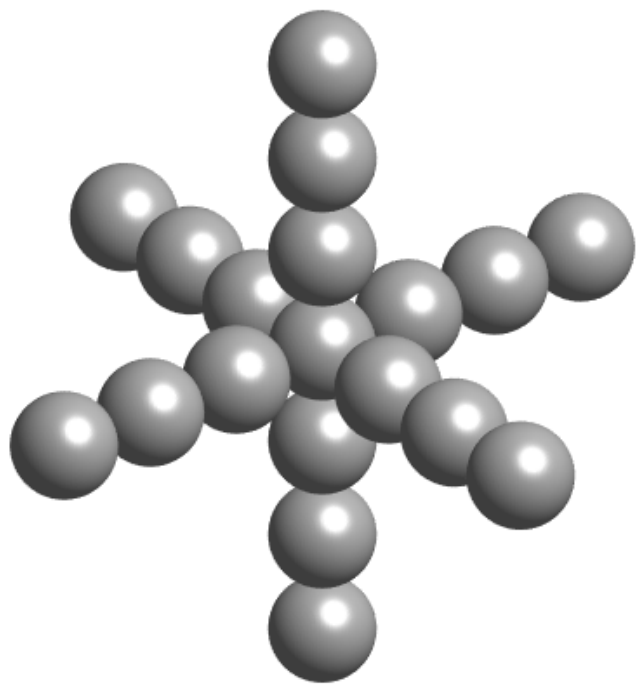}
    \includegraphics[width=0.59\textwidth]{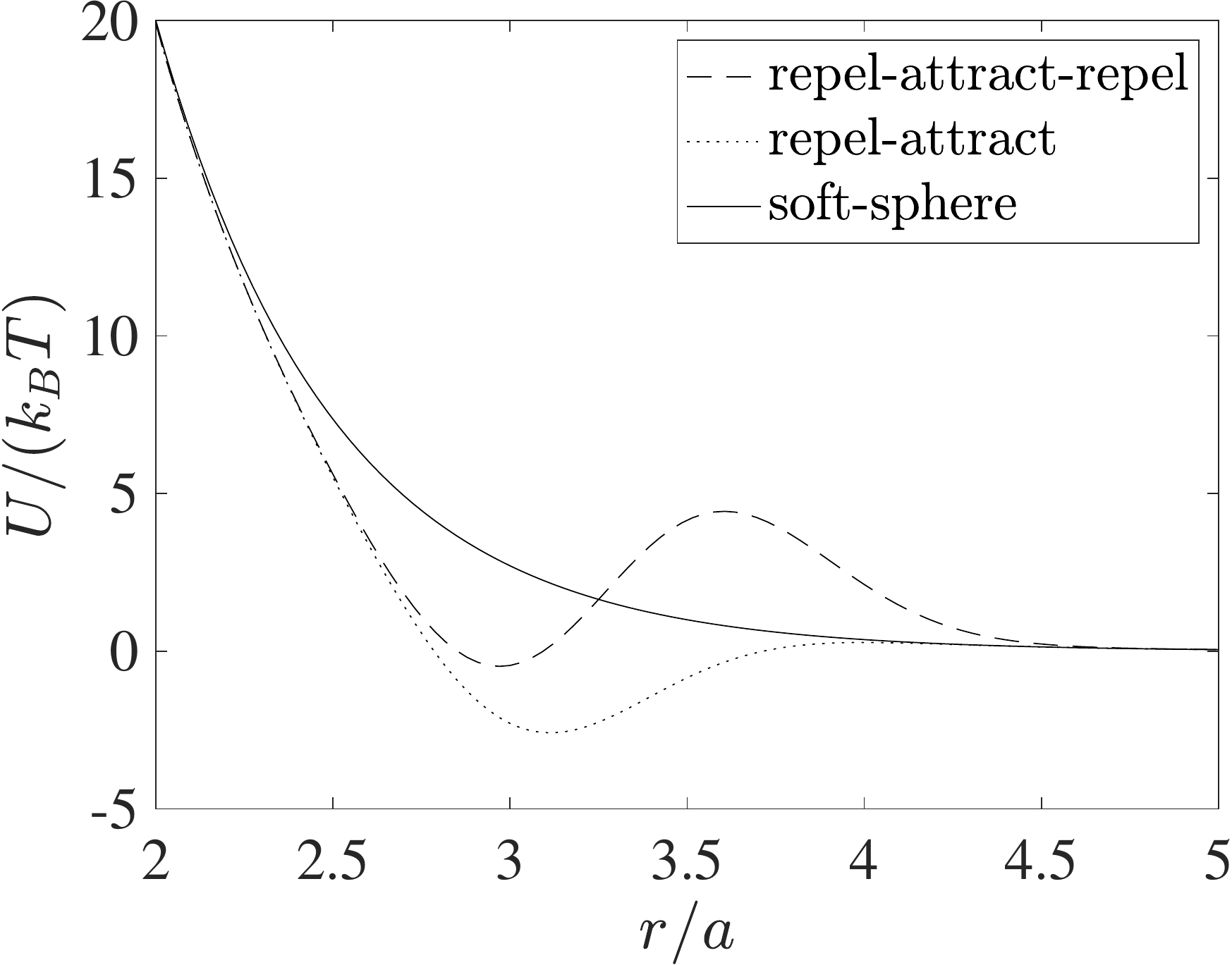}
    \caption{An illustration of a CH particle formed of 19 FCM-particles of radius $a$ (left), and the different interaction potentials for suspended CH particles (right).}
    \label{fig:interaction_potentials}
    \label{fig:star_shaped_particle}
\end{figure}

In this section, we use fluctuating FCM with the gDC to perform nonequilibrium simulations that examine the rheology of Brownian Czech hedgehog-shaped (CH) particles, depicted in \cref{fig:star_shaped_particle}.  This set of simulations is inspired by the recent experimental work of Bourrianne \textit{et al.} \cite{bourrianne_unifying_2020}, where suspensions of dendritic, silica particles, either hydrophobic or hydrophilic, were found to exhibit interesting rheological behaviour, including discontinuous shear thickening, depending on the particle interactions, as well as the relative strengths of shear and particle diffusion (Peclet number) and volume fraction.  

In our simulations, the CH particles are formed of 19 FCM-particles of radius $a$, as shown in \cref{fig:star_shaped_particle}. This shape was chosen to reproduce, at some level, the complex structure and high specific surface area of the silica particles in the experiments.  We consider two types of interactions between the CH particles.  For the first type, the FCM-particles comprising the CH particles repel at short-range through a soft potential, while attracting at longer range.  Specifically, defining
\begin{equation}
    \phi_{r_c}\left(r\right) = A\exp\left(-\left(\frac{r - r_c}{\lambda}\right)^2\right)
\end{equation}
for $A = 5k_BT$ and $\lambda = 0.5a$, the `repel-attract' potential is given by 
\begin{align}
    U_1(r) &= U(r) - \phi_{3a}(r),
\end{align}
where $U(r)$ is the soft-sphere potential defined in \cref{eqn:soft_sphere_potential}. For the second type of interaction, we have the `repel-attract-repel' potential
\begin{align}
    U_2(r) &= U(r) - \phi_{3a}(r) + \phi_{7a/2}(r),
\end{align}
where a barrier is introduced in the potential just beyond the well.  The resulting shapes of these potentials are shown in \cref{fig:interaction_potentials}. Note that strictly these definitions are only $r \geq 2a$, with the potentials continuously extended by linear functions for $r < 2a$ to ensure that the force is constant for blobs which overlap, just as for the original soft-sphere potential $U(r)$.
%
%
%


The simulations are performed in a triply-periodic computational domain with dimensions $L_x = L_y = L_z = 77.65a$.  In order ensure that a strong rheological response is observed, a high volume fraction of 20\% was used for all simulations.  This corresponds to 1177 CH particles, yielding a mobility problem for the $1177 \times 19 = 22363$ FCM-particles in each simulation. Simulations were run with a time step length of $\Delta t = \tau_D/600$.  The initial conditions for the full simulations were constructed by running simulations that ignore thermal fluctuations and hydrodynamic interactions, i.e. using a diagonal mobility matrix, but retain interactions due to the potentials.  In practice, we find that by allowing the CH particles to settle in this way before the shear is applied, the `long-time' velocity profile is realised more quickly.  The simulations were run to final times between $35\tau_D/3$ and $50\tau_D$ and in all cases, a regular oscillatory velocity profile emerges long before the end of the simulation.  

To measure the suspension viscosity from this emergent velocity field, we adopt the approach described by V\'azquez \textit{et al.} \cite{vazquez-quesada_multiblob_2014}.  Here, the background fluid is forced using the periodic force density
\begin{equation}
    \v{f}(x,y,z) = f_0\sin\left(k y\right)\uv{e}_x
\end{equation}
where $f_0$ is the force magnitude and $k = 2\pi /L_y$ is the wavenumber.  The $x$-component of the resulting velocity field of the suspension is averaged over time and in the $x$- and $z$-directions, resulting in a sinusoidal velocity profile in $y$ with magnitude $v_x$. The effective viscosity of the suspension can then be defined as
\begin{equation}
    \eta_{eff} = \frac{f_0}{k^2 v_x}.
\end{equation}

\begin{figure}[htb!]
    \centering
    \includegraphics[width=0.51\textwidth]{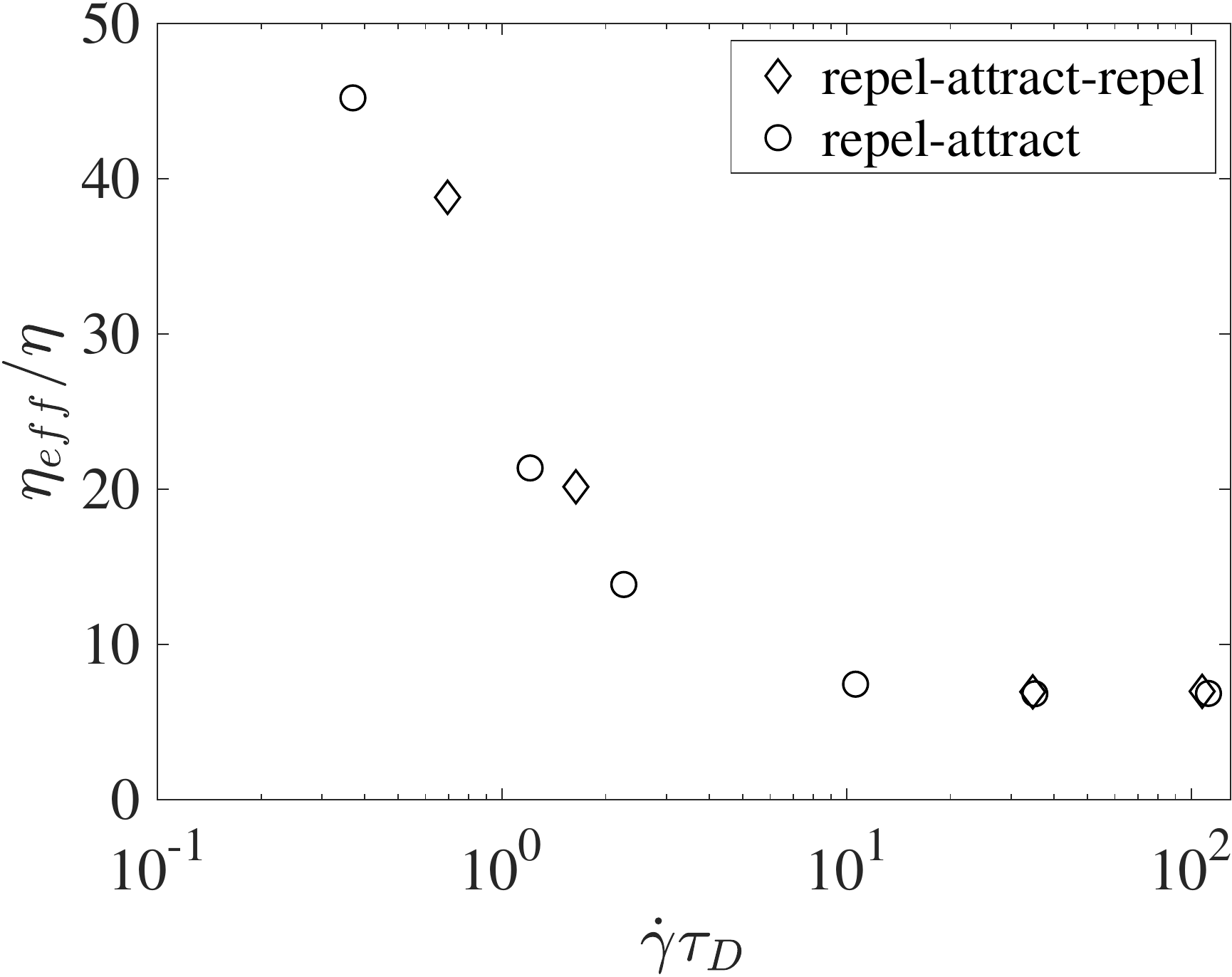}
    \includegraphics[width=0.49\textwidth]{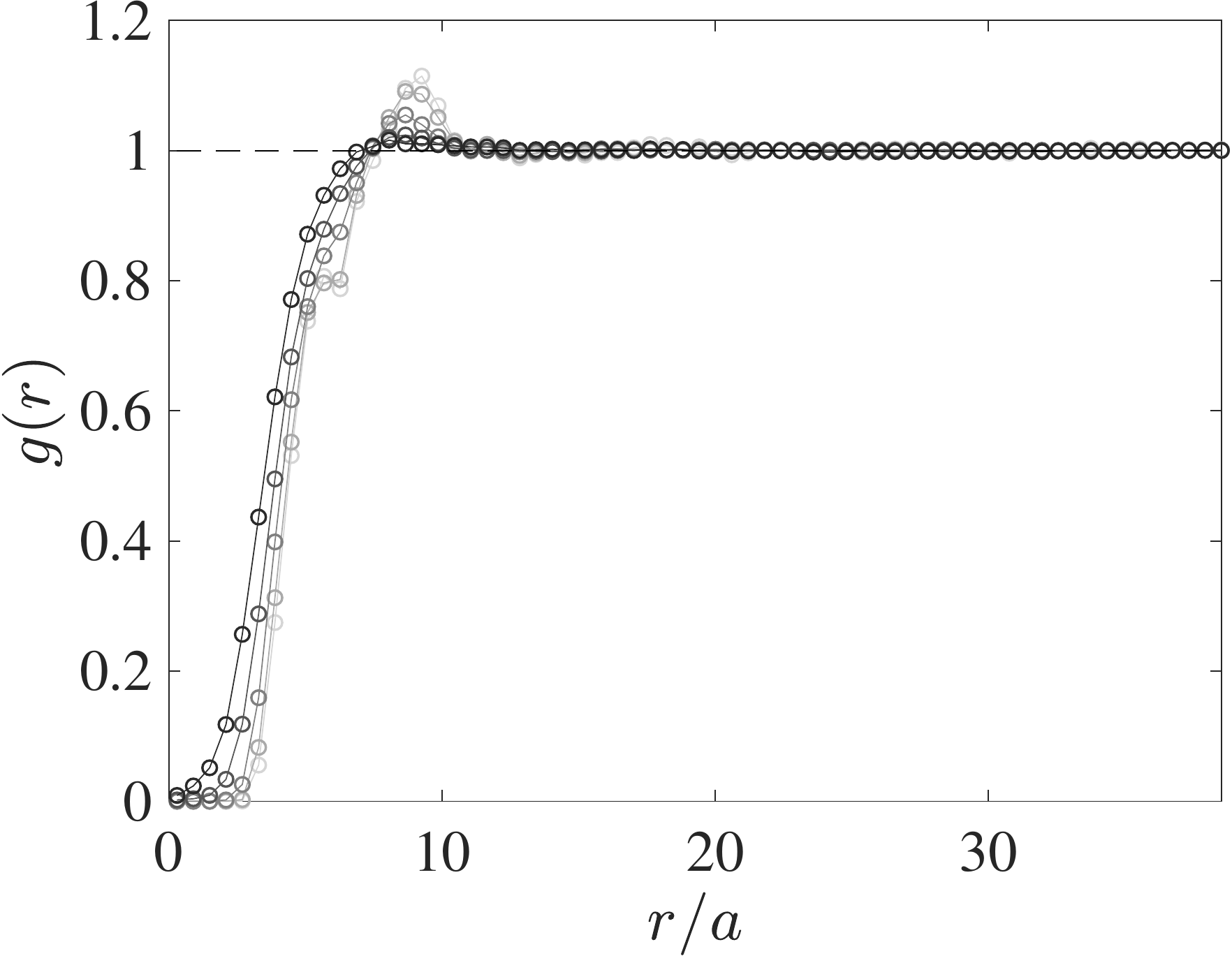}
    \def\svgwidth{0.49\columnwidth}
    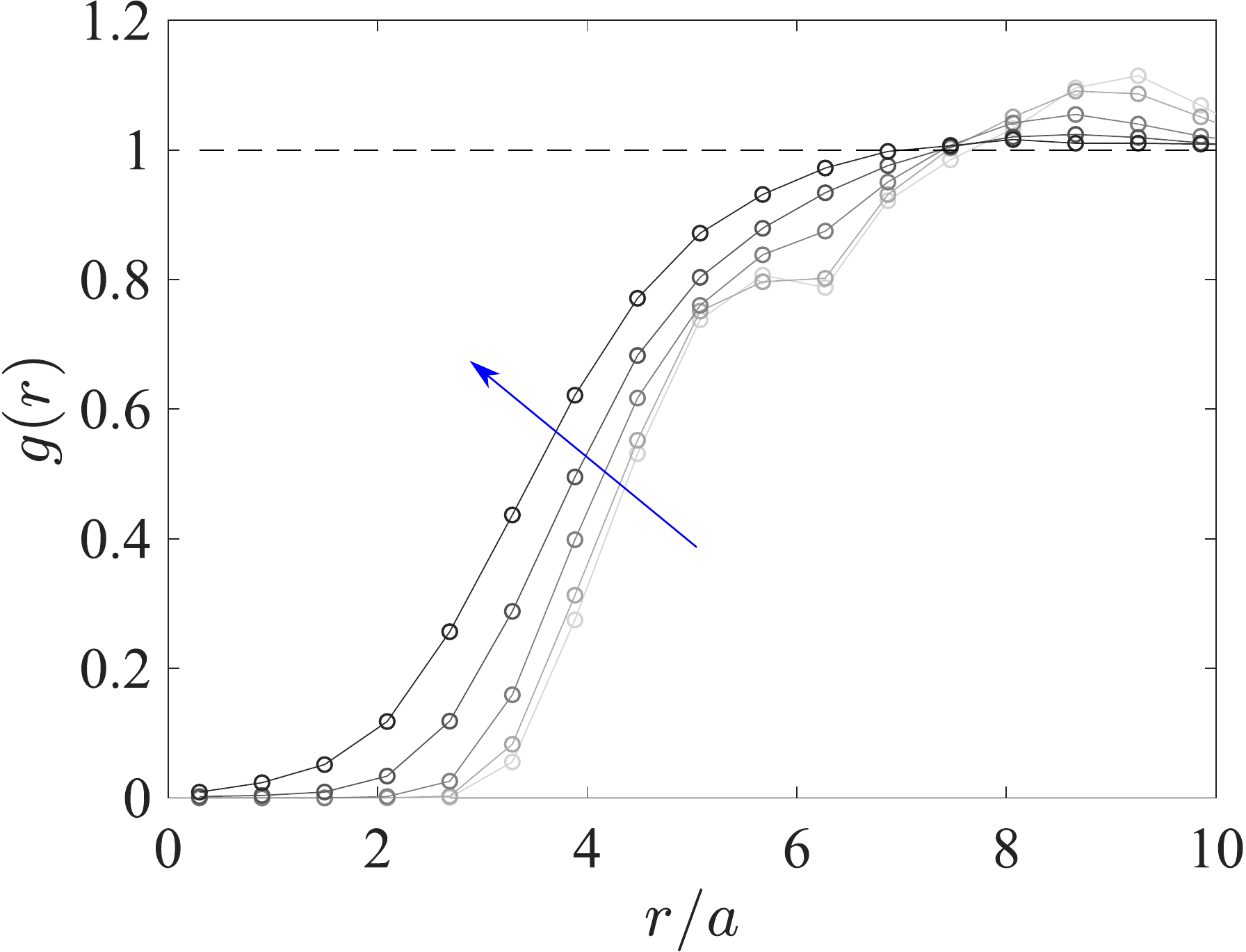
    \caption{The effective viscosity of both types of suspension for different dimensionless shear rates (top), and the radial distribution functions for the different `repel-attract' simulations (bottom). The bottom right plot is displaying the same data as the bottom left plot but over a smaller $r$ range. The colour of the curves fades from black to white as $f_0$ decreases.}
    \label{fig:hydrophobic_radial_distribution}
    \label{fig:visc_vs_shear_plot}
\end{figure}

In \cref{fig:visc_vs_shear_plot}, the ratio of the effective viscosity $\eta_{eff}$ to the underlying fluid viscosity $\eta$ is compared to the shear rate $\dot{\gamma}$ for suspensions with both types of particle interactions.  In this work, $\dot{\gamma}$ is taken as the maximum slope of the sinusoidal velocity profile.  We find that both the `repel-attract' and `repel-attract-repel' exhibit rapid shear thinning at lower shear rates, as found in the original experiments of Bourrianne \textit{et al.}  We do not observe, however, any shear thickening for the `repel-attract-repel' CH particle suspensions whose interactions are meant to be similar to those of the hydrophilic particles in the experiments.  Bourrianne \textit{et al.} suggest that solid friction is required for the onset of shear thickening and that hydrogen bonding is additionally required for DST to be observed.  Both of these interactions are absent from our simulations.

Given that this is a non-equilibrium problem, it is of particular interest to examine distributions of the particle states since they cannot be sampled using equilibrium procedures such as MCMC. \Cref{fig:hydrophobic_radial_distribution} shows the radial distribution function $g(r)$ where $r$ is the centre-to-centre distance of $r$ for the CH particles with `repel-attract' interactions; the distribution for `repel-attract-repel' interactions is similar.  Simulations with higher values of $f_0$ correspond to darker curves in the figure.  At the smallest $f_0$, where the particle motion is dominated by the Brownian motion, $g(r)$ increases from zero to a peak value at separations between $r/a \approx 8$ and $r/a \approx 10$.  At larger separations, $g(r)$ gradually decreases to approach values for a uniform distribution.  As $f_0$, and hence the shear rate, is increased, this peak diminishes and the distribution approaches a monotonically increasing function.


\section{Summary and conclusions}

In this paper, we developed the gDC scheme for the efficient time integration of hydrodynamically interacting rigid Brownian particles.  The gDC automatically accounts for Brownian drift in advancing particle positions and orientations while retaining the need to perform only a single, full mobility problem at each time step, while achieving accuracy comparable to the current state-of-the-art.  The gDC has been designed to be used with fast methods for applying the mobility matrix and generating random increments with the correct covariance and is ideally suited for grid-based computations that take advantage of fluctuating hydrodynamics.  Additionally, many of the additional computations that the gDC requires are local to each particle, enabling parallelisation and large-scale computation.  Our simulations demonstrate each of these features, as well as the applicability of gDC to enable simulations that complement experiments in modern suspension mechanics and colloidal science.  

\section*{Acknowledgements}
EEK and TAW gratefully acknowledge support from EPSRC Grant EP/P013651/1. TAW is also thankful for funding through an EPSRC Studentship (Ref: 1832024). BD acknowledges support from the French National Research Agency (ANR), under award ANR-20-CE30-0006.

\appendix
\section{Quaternions}
\label{sec:quaternions}

In order to describe the state of an arbitrarily shaped particle, its orientation must be provided along with its position in space. This requires providing the rotation mapping from some reference configuration to the current orientation of the particle and thus a parametrisation for three dimensional rotations must be chosen. Perhaps the most natural choice is the axis-angle representation $\v{\theta} = \theta\uv{e}$, which describes the anti-clockwise rotation through an angle $\theta$ about the axis $\uv{e}$, but subsequent rotations cannot be easily combined under this representation. This problem is shared by the Euler angles, which further suffer from the need to specify a convention for their application as well as the gimbal lock phenomenon, wherein certain choices of one angle result in a loss of a degree of freedom. These issues are avoided when using rotation matrices, but they require significant amounts of redundant storage -- storing nine variables to represent three degrees of freedom -- and they must be projected back to the nearest orthonormal matrix in the event of accumulated numerical error. An alternative representation, and the one used in this work, is that of the unit quaternions.

The unit quaternions can be identified with the unit sphere in $\Reals^4$, although they are often viewed as having separate scalar and vector parts and thus written as $\q{q} = (q_0, \v{q})$, where $q_0 \in \Reals$ and $\v{q} \in \Reals^3$; the unit length constraint is then expressed as $q_0^2 + \|\v{q}\|^2 = 1$. They form a group under the non-commuting, associative multiplication known as the Hamilton product,
\begin{equation}\begin{split}
    \q{p}\qprod\q{q} &= (p_0, \v{p}) \qprod (q_0, \v{q}) \\ &= (p_0q_0 - \v{p}\cdot\v{q}, p_0\v{q} + q_0\v{p} + \v{p}\times\v{q}),
\end{split}\end{equation}
with inverses $\q{q}^{-1} = (q_0, -\v{q})$ and identity $\q{I}_q = (1,\vu{0})$. It is through the Hamilton product and the polar form for unit quaternions,
\begin{equation} \label{eqn:quaternion_polar_form}
    \q{q} = \left(\cos\left(\frac{\theta}{2}\right),\;\sin\left(\frac{\theta}{2}\right)\uv{e}\right),
\end{equation}
that the unit quaternions can be associated with the anti-clockwise rotation through an angle $\theta$ about the axis $\uv{e}$, with the image, $\v{x}'$, of a vector $\v{x}$ under this rotation satisfying
\begin{equation}
    (0, \v{x}') = \q{q} \qprod (0, \v{x}) \qprod \q{q}^{-1}.
\end{equation}
Thus it is clear that the Hamilton product allows us to describe successive quaternion-represented rotations as another unit quaternion; rotating according to $\q{p}\qprod\q{q}$ is equivalent to first rotating according to $\q{q}$ before rotating according to $\q{p}$. Consequently, the unit quaternions improve on the rotation matrix representation of rotations by reducing the storage cost from nine to four variables, reducing the cost of correcting accumulated errors to normalising a vector and by providing an intuitive correspondence with the axis-angle description.

If $\v{\Omega}$ denotes the angular velocity of the body, then the unit quaternion associated with the orientation of the body, viewed as an element of $\Reals^4$, satisfies
\begin{equation}
    \fd{\q{q}}{t} = \t{\Psi}_\q{q}\v{\Omega},
\end{equation}
where
\begin{equation} \label{eqn:psi_mat_def}
    \t{\Psi}_\q{q} = \frac{1}{2}\begin{bmatrix}-\v{q}^\top \\ q_0\t{I} + \left[\times\v{q}\right]\end{bmatrix} \in \Reals^{4\times 3}
\end{equation}
satisfies $\t{\Psi}_\q{q}\v{\Omega} = \frac{1}{2}\left(0,\v{\Omega}\right)\qprod\q{q}$.

Given an initial orientation quaternion $\q{q}_0$ at time $t = 0$, this ODE could be solved using standard linear techniques. However, this will produce a quaternion with non-unit norm in general. The resulting quaternion could be projected back to unit length, but this represents an unnecessary source of error. Instead, we utilise the Lie algebra associated with the group of unit quaternions to perform multiplicative updates which preserve the norm to machine precision. Following the general frameworks of Iserles \textit{et al.} \cite{iserles_lie-group_2000} and Faltinsen \textit{et al.} \cite{faltinsen_multistep_2001}, we observe that the solution of the ODE for the orientation quaternion,
    \begin{equation} \begin{split}
        \fd{\q{q}}{t} &= \t{\Psi}_\q{q}\v{\Omega},\\
        \q{q}(0) &= \q{q}_0,
    \end{split}\end{equation}
    is given by
    \begin{equation} \label{eqn:quaternion_from_lie}
        \q{q}(t) = \exp\left(\v{u}(t)\right)\qprod\q{q}_0,
    \end{equation}
    where $\v{u}$ is the so-called Lie algebra element satisfying
    \begin{equation} \begin{split} \label{eqn:lie_ode}
        \fd{\v{u}}{t} &= \t{D}^{-1}_\v{u}\v{\Omega},\\
        \v{u}(0) &= 0.
    \end{split}\end{equation}
    The exponential function for this Lie algebra corresponds to the polar form of the unit quaternions, i.e.\
    \begin{equation}
        \exp\left(\v{u}\right) = \left(\cos\left(\frac{\|\v{u}\|}{2}\right),\;\sin\left(\frac{\|\v{u}\|}{2}\right)\frac{\v{u}}{\|\v{u}\|}\right),
    \end{equation}
    and $\t{D}^{-1}_\v{u}$ is the `dexpinv' matrix defined by
    \begin{equation} \label{eqn:dexpinv2}
        \t{D}^{-1}_\v{u} = \t{I} - \frac{1}{2}\left[\v{u}\times\right] - \frac{1}{2\|\v{u}\|^2}\left(\|\v{u}\|\cot\left(\frac{\|\v{u}\|}{2}\right) - 2\right)\left[\v{u}\times\right]^2;
    \end{equation}
    the Lie algebra considered here is isomorphic to $(\Reals^3, \times)$ and so $\v{u}$ should simply be thought of as a 3-vector. It should also be noted that $\t{D}^{-1}_\v{u}$ is continuously extended by $\t{D}^{-1}_\vu{0} = \t{I}$.
    
    As discussed by Faltinsen \textit{et al.} \cite{faltinsen_multistep_2001}, the numerical properties of the solution are preserved by in fact using $\v{u}$ to describe the rotation we apply to the current quaternion rather than the initial condition $\q{q}_0$. That is, at a given discrete time $n$ (i.e.\ $t = t^n = n\Delta t$) we solve for $\v{u}^{n+1}$ assuming that $\v{u}^n = \vu{0}$ and define $\q{q}^{n+1} = \exp(\v{u}^{n+1})\qprod\q{q}^n$. The fact that the Lie algebra element is always expressed in a coordinate system such that it is zero at the start of the current time-step will allow us to make some useful simplifications in \cref{sec:consistency}.
    
\section{Expressing torques in terms of the Lie algebra element}
\label{sec:lie_torque}

Delong \textit{et al.} \cite{delong_brownian_2015} considered the change in energy due to an infinitesimal rotation to argue that if the quaternion $\q{q}$ describes the orientation of a particle then the torque it experiences due to a potential $U$ can be written
\begin{equation} \label{eqn:quaternion_torque}
    \v{T} = -\t{\Psi}_\q{q}^\top\p_\q{q}U,
\end{equation}
where the matrix $\t{\Psi}_\q{q}$ is as defined in \cref{eqn:psi_mat_def}. In this section, we employ a similar argument to show that the analogous result holds for the corresponding Lie algebra element.

Consider the torque $\v{T} = -\partial_\v{\varphi}U$ generated by the potential $U$, where $\v{\varphi}$ is the oriented angle. We want to express this torque directly in terms of the Lie algebra element $\v{u}$. Recalling \cref{eqn:lie_ode}, we have
\begin{equation}\begin{split}
    d\v{u} &= \t{D}^{-1}_\v{u}d\v{\varphi} \\ &= d\v{\varphi} - \frac{1}{2}\v{u}\times d\v{\varphi} - g\left(\v{u}\right)\v{u}\times\left(\v{u}\times d\v{\varphi}\right),
\end{split}\end{equation}
where we have defined $g(\v{u}) = \left(\|\v{u}\|\cot\left(\frac{\|\v{u}\|}{2}\right) - 2\right)/\left(2\|\v{u}\|^2\right)$ for legibility. Hence, the change in energy $dU$ due to an infinitesimal rotation $d\v{\varphi}$ satisfies
\begin{equation}\begin{split}
    -\v{T}\cdot d\v{\varphi} = dU &= \partial_\v{u}U\cdot d\v{u} \\ &= \partial_\v{u}U\cdot d\v{\varphi} - \frac{1}{2}\left(\v{u}\times d\v{\varphi}\right)\cdot\partial_\v{u}U - g\left(\v{u}\right)\left[\v{u}\times\left(\v{u}\times d\v{\varphi}\right)\right]\cdot\partial_\v{u}U.
\end{split}\end{equation}
Since the scalar triple product is invariant under cyclic permutation of the vectors we observe that $\left(\v{u}\times d\v{\varphi}\right)\cdot\partial_\v{u}U = \left(\partial_\v{u}U\times \v{u}\right)\cdot d\v{\varphi}$ and $\left[\v{u}\times\left(\v{u}\times d\v{\varphi}\right)\right]\cdot\partial_\v{u}U = \left[\partial_\v{u}U\times\v{u}\right]\cdot\left(\v{u}\times d\v{\varphi}\right) = \left[\left(\partial_\v{u}U\times\v{u}\right)\times\v{u}\right]\cdot d\v{\varphi}$, giving us
\begin{equation}
    dU = \left[\partial_\v{u}U - \frac{1}{2}\left(\partial_\v{u}U\times \v{u}\right) - g\left(\v{u}\right)\left(\partial_\v{u}U\times\v{u}\right)\times\v{u}\right]\cdot d\v{\varphi}.
\end{equation}
This permits the identification of the torque as
\begin{equation}\begin{split}
    \v{T} &= -\left[\partial_\v{u}U - \frac{1}{2}\left(\partial_\v{u}U\times \v{u}\right) - g\left(\v{u}\right)\left(\partial_\v{u}U\times\v{u}\right)\times\v{u}\right] \\ &= -\left[\t{I} - \frac{1}{2}\left[\times \v{u}\right] - g\left(\v{u}\right)\left[\times\v{u}\right]^2 \right]\partial_\v{u}U \\ &= -\t{D}^{-\top}_\v{u}\partial_\v{u}U.
\end{split}\end{equation}

\section{Equivalence between the Lie algebra and quaternion SDEs}
\label{sec:quaternion_sde}

Having shown that the torque on a body can be written in terms of the unit quaternion describing its orientation, $\q{q}$, according to \cref{eqn:quaternion_torque}, Delong \textit{et al.} \cite{delong_brownian_2015} argue that $\q{q}$ should satisfy the It\^o Langevin SDE
\begin{equation} \label{eqn:quaternion_sde}
    d\q{q} = \left(k_BT\partial_\q{q}\cdot\left(\t{\Psi}_\q{q}\t{N}_{\Omega T}\t{\Psi}_\q{q}^\top\right) - \t{\Psi}_\q{q}\t{N}_{\Omega T}\t{\Psi}_\q{q}^\top\partial_\q{q}U\right)dt + \sqrt{2k_BT}\t{\Psi}_\q{q}\t{N}_{\Omega T}^{1/2}d\v{W},
\end{equation}
where $\t{N}_{\Omega T}$ is the rotational mobility matrix mapping the torque on the body, $\v{T} = -\t{\Psi}_\q{q}^\top\p_\q{q}U$, to its angular velocity, $\v{\Omega}$, and $\t{N}_{\Omega T}^{1/2}$ satisfies $\t{N}_{\Omega T}^{1/2}\left(\t{N}_{\Omega T}^{1/2}\right)^\top = \t{N}_{\Omega T}$ as required by the fluctuation-dissipation theorem. In this work, we integrate unit quaternions in time using their associated Lie algebra elements rather than handle the quaternions directly. Thus, we seek an It\^o SDE for the Lie algebra element such that the resulting quaternion, defined by \cref{eqn:quaternion_from_lie}, satisfies \cref{eqn:quaternion_sde} and hence that the rotational dynamics in our approach are equivalent to those in the work of Delong \textit{et al.} \cite{delong_brownian_2015}. Given our result in \cref{sec:lie_torque}, namely that the torque on the body can be expressed using the Lie algebra element as $\v{T} = -\t{D}^{-\top}_\v{u}\partial_\v{u}U$, we propose that the appropriate It\^o SDE is
\begin{equation} \label{eqn:lie_ito_sde}
    d\v{u} = \left(k_BT\partial_\v{u}\cdot\left(\t{D}^{-1}_\v{u}\t{N}_{\Omega T}\t{D}^{-\top}_\v{u}\right) - \t{D}^{-1}_\v{u}\t{N}_{\Omega T}\t{D}^{-\top}_\v{u}\partial_\v{u}U\right)dt + \sqrt{2k_BT}\t{D}^{-1}_\v{u}\t{N}_{\Omega T}^{1/2}d\v{W},
\end{equation}
the direct analogue of \cref{eqn:quaternion_sde}. In the remainder of this section, we demonstrate that this does indeed result in the correct It\^o SDE for the unit quaternion.

Loosely speaking, the approach here is to convert the It\^o SDE for $\v{u}$ to the Stratonovich interpretation, for which many standard calculus results are recovered, then to mirror the deterministic result ``$\fd{\v{u}}{t} = \t{D}^{-1}_\v{u}\v{\Omega} \implies \fd{\q{q}}{t} = \t{\Psi}_\q{q}\v{\Omega}$'' and finally to recover the It\^o interpretation. To that end, recall that if $\v{y}$ satisfies the It\^o SDE
\begin{equation} \label{eqn:generic_ito_sde}
    d\v{y} = \v{h}\left(\v{y}\right)dt + \t{\gamma}\left(\v{y}\right)d\v{W},
\end{equation}
then the equivalent Stratonovich form of the SDE is
\begin{equation}
    d\v{y} = \left(\v{h}\left(\v{y}\right) - \frac{1}{2}\v{c}\left(\v{y}\right)\right)dt + \t{\gamma}\left(\v{y}\right)\circ d\v{W},
\end{equation}
where $c_i = \gamma_{jk}\pd{\gamma_{ik}}{y_j}$ and $\circ$ denotes the Stratonovitch product. Writing $\t{D}^{-1} = \t{D}^{-1}_\v{u}$ and $\t{N} = \t{N}_{\Omega T}$ (and hence $\t{N}^{1/2} = \t{N}_{\Omega T}^{1/2}$) to simplify notation, we make the identification $\gamma_{ij} = \sqrt{2k_BT}D^{-1}_{ik}N^{1/2}_{kj}$ by comparing \cref{eqn:lie_ito_sde} to \cref{eqn:generic_ito_sde} for $\v{y} = \v{u}$ and hence obtain
\begin{equation}\begin{split}
    \frac{1}{2}c_i &= k_B T D^{-1}_{jb}N^{1/2}_{bk}\pd{\left(D^{-1}_{ia}N^{1/2}_{ak}\right)}{u_j} \\ &= k_B T \left(\pd{\left(D^{-1}_{ia}N^{1/2}_{ak}N^{1/2}_{kb}D^{-1}_{jb}\right)}{u_j} - D^{-1}_{ia}N^{1/2}_{ak}\pd{\left(D^{-1}_{jb}N^{1/2}_{bk}\right)}{u_j}\right) \\ &= k_B T \left(\pd{\tilde{N}_{ij}}{u_j} - D^{-1}_{ia}N^{1/2}_{ak}\pd{\left(D^{-1}_{jb}N^{1/2}_{bk}\right)}{u_j}\right).
\end{split}\end{equation}
Now we claim that $\pd{D^{-1}_{jb}}{u_j} = 0$ and hence that $\pd{\left(D^{-1}_{jb}N^{1/2}_{bk}\right)}{u_j} = D^{-1}_{jb}\pd{N^{1/2}_{bk}}{u_j}$. For $\v{u} \neq \vu{0}$, we can simply differentiate \cref{eqn:dexpinv} and observe that $\pd{D^{-1}_{jb}}{u_j} \propto \left(\left[\v{u}\times\right]^2\v{u}\right)_b = 0$. The case $\v{u} = \vu{0}$ can be shown easily using the limit definition of the partial derivative. Thus
\begin{equation}
    \frac{1}{2}c_i = k_B T \left(\pd{\tilde{N}_{ij}}{u_j} - D^{-1}_{ia}N^{1/2}_{ak}D^{-1}_{jb}\pd{N^{1/2}_{bk}}{u_j}\right),
\end{equation}
which allows us to cast \cref{eqn:lie_ito_sde} in Stratonovich form as
\begin{equation} \label{eqn:lie_strat_sde} \begin{split}
    du_i &= \left(k_B T D^{-1}_{ia}N^{1/2}_{ak}D^{-1}_{jb}\pd{N^{1/2}_{bk}}{u_j} - \tilde{N}_{ij}\pd{U}{u_j}\right)dt + \sqrt{2k_BT}D^{-1}_{ia}N^{1/2}_{aj}\circ dW_j\\ &= D^{-1}_{ia}\left[\left(k_B T N^{1/2}_{ak}D^{-1}_{jb}\pd{N^{1/2}_{bk}}{u_j} - N_{ab}D^{-1}_{jb}\pd{U}{u_j}\right)dt + \sqrt{2k_BT}N^{1/2}_{aj}\circ dW_j\right].
\end{split}\end{equation}
Applying Theorem 5.1 of Malham and Wiese \cite{Malham_2008} (recognizing the $\v{\xi}$ as the columns of $\t{N}^{1/2}$) we find that the corresponding orientation quaternion satisfies the Stratonovich equation
\begin{equation}
    dq_i = \Psi_{ia}\left[\left(k_B T N^{1/2}_{ak}D^{-1}_{jb}\pd{N^{1/2}_{bk}}{u_j} - N_{ab}D^{-1}_{jb}\pd{U}{u_j}\right)dt + \sqrt{2k_BT}N^{1/2}_{aj}\circ dW_j\right],
\end{equation}
where we have dropped the $\q{q}$ subscript from $\t{\Psi}_\q{q}$ for legibility. Recalling \cref{sec:lie_torque}, we make the identification $D^{-1}_{jb}\pd{}{u_j} = \Psi_{jb}\pd{}{q_j}$ and thus the above becomes
\begin{equation}
    dq_i = \Psi_{ia}\left[\left(k_B T N^{1/2}_{ak}\Psi_{jb}\pd{N^{1/2}_{bk}}{q_j} - N_{ab}\Psi_{jb}\pd{U}{q_j}\right)dt + \sqrt{2k_BT}N^{1/2}_{aj}\circ dW_j\right].
\end{equation}
Observing that the form of this equation is analogous to \cref{eqn:lie_strat_sde} and noting that $\pd{\Psi_{jb}}{q_j} = 0$ since the $j^\mathrm{th}$ row of $\t{\Psi}$ doesn't depend on $q_j$, we simply perform the above procedure in reverse to obtain \cref{eqn:quaternion_sde}.

Having established the appropriate It\^o SDE for the Lie algebra element, it is simple to couple this to the corresponding equation for the position of the body to produce the complete It\^o SDE for the state of the body we presented in \cref{eqn:system_sde}.

\section{Algorithm for Euler-Maruyama scheme}
 \label{sec:EM_alg}
For each time-step $n = 0,1,2,...$
\begin{enumerate}

    \item Generate a vector  $\v{W}$ (or tensor $\t{W}$ if using fluctuating hydrodynamics) of $\mathcal{N}(0,1)$ random variables.
 
    \item Solve the full problem
    \begin{equation}
        \begin{bmatrix}
        \t{M}^n & -\t{K}^n \\
        -\left(\t{K}^\top\right)^n  & \vu{0}
        \end{bmatrix}
        \begin{bmatrix}
        \v{\lambda}^n \\ \v{V}^n
        \end{bmatrix}
         = \begin{bmatrix}
         -\breve{\v{v}}^n = -\sqrt{\frac{2 k_B T}{\Delta t}}\left(\t{M}^{1/2}\right)^n\v{W} \\ -\v{F}^n
         \end{bmatrix}
    \end{equation}
    for
     \begin{equation}
    \v{V}^n = \t{N}^n\left(\v{F}^n + \left(\t{K}^\top\right)^n\left(\t{M}^{-1}\right)^n\breve{\v{v}}^n\right) = \t{N}^n\v{F}^n + \breve{\v{V}}^n ,
    \end{equation}
    
    \item Move the rigid bodies to the next time-step according to
    \begin{align}
        \v{x}^{n+1} &= \v{x}^n + \Delta t \v{V}^n, \\
        \q{q}^{n+1}_p &= \exp\left(\v{u}^{n+1}_p\right)\qprod\q{q}^n_p \text{ for } 1 \leq p \leq N.
    \end{align}
    
\end{enumerate}

\section{Proof of consistency for GDC}
\label{sec:consistency}

In this section we offer a proof of the consistency of the gDC algorithm (\cref{sec:gDC_alg}); i.e. that the increment $\Delta \v{x}^n = \v{x}^{n+1} - \v{x}^n = \nu\Delta t\v{V}^m$ produces the correct first and second moments for freely-diffusing rigid bodies,
\begin{align}
    \label{eqn:old_first_moment} \langle\Delta\v{x}^n\rangle &= k_BT\Delta t\left(\p_\v{x}\cdot\tilde{\t{N}}\right)^n, \\ \label{eqn:old_second_moment} \langle\Delta\v{x}^n\left(\Delta\v{x}^n\right)^\top\rangle &= 2k_BT\Delta t\tilde{\t{N}}^n,
\end{align}
to first-order in time. We begin by observing that $\t{D}^{-1}_\vu{0} = \t{I} \implies \t{\Phi}^n = \t{I}$ (recall \cref{sec:quaternions}) and hence that \cref{eqn:old_second_moment} reduces to
\begin{equation}
    \label{eqn:second_moment} \langle\Delta\v{x}^n\left(\Delta\v{x}^n\right)^\top\rangle = 2k_BT\Delta t\t{N}^n.
\end{equation}

To similarly simplify \cref{eqn:old_first_moment} we introduce the shorthand $\partial_i \equiv \left(\partial_\v{x}\right)_i = \partial/\partial x_i$ and, summing over repeated indices, expand
\begin{equation}\begin{split}
    \left(\p_j\tilde{N}_{ij}\right)^n &= \left(\p_j\left(\Phi_{ia}N_{ab}\Phi_{jb}\right)\right)^n \\ &= \Phi_{ia}^nN_{ab}^n\left(\p_j\Phi_{jb}\right)^n + \Phi_{ia}^n\Phi_{jb}^n\left(\p_jN_{ab}\right)^n + N_{ab}^n\Phi_{jb}^n\left(\p_j\Phi_{ia}\right)^n \\ &= N_{ib}^n\left(\p_j\Phi_{jb}\right)^n + \left(\p_jN_{ij}\right)^n + N_{aj}^n\left(\p_j\Phi_{ia}\right)^n,
\end{split}\end{equation}
since $\Phi^n_{kl} = \delta_{kl}$. In \cref{sec:quaternion_sde} we show that $\partial_j\left(\t{D}^{-1}_\v{u}\right)_{jb} = 0$ and hence the first term vanishes. Furthermore, applying the definition of partial differentiation yields $\left(\partial_j\left(\t{D}^{-1}_\v{u}\right)_{ia}\right)^n = -\frac{1}{2}\epsilon_{ija}$, and hence the components of the third term which don't vanish in the differentiation vanish as the inner product of a symmetric and skew-symmetric matrix. Thus we obtain the simpler expression
\begin{equation}
    \label{eqn:first_moment} \langle\Delta\v{x}^n\rangle = k_BT\Delta t\left(\partial_\v{x}\cdot\t{N}\right)^n
\end{equation}
for the first moment in expectation.

To simplify notation, for the remainder of this section all quantities are evaluated at time $n$ unless explicitly indicated otherwise. We start by expanding $\v{V}^m$ about time $t_n$ to obtain
\begin{equation}
    V^m_i = V_i + \frac{\Delta t}{2}\mathcal{V}_j\p_jV_i + \frac{\Delta t^2}{8}\mathcal{V}_k\mathcal{V}_j\p_k\p_jV_i + \order(\Delta t).
\end{equation}
Recalling that $\nu = 1 + \frac{\Delta t}{2}\p_j\mathcal{V}_j$ we obtain
\begin{equation}\label{eqn:consistency_full_rhs}\begin{split}
    \nu V^m_i = V_i &+ \frac{\Delta t}{2}\mathcal{V}_j\p_jV_i + \frac{\Delta t^2}{8}\mathcal{V}_k\mathcal{V}_j\p_k\p_jV_i \\ &+ \frac{\Delta t}{2}V_i\p_j\mathcal{V}_j + \frac{\Delta t^2}{4}\mathcal{V}_j\p_j\left(V_i\right)\p_k\left(\mathcal{V}_k\right) + \order(\Delta t).
\end{split}\end{equation}
We recognize the second and fourth terms on the right hand side as a product rule expansion and re-write the above as
\begin{equation}\begin{split}\label{eqn:consistency_simplified_rhs}
    \nu V^m_i = V_i &+ \frac{\Delta t}{2}\p_j\left(\mathcal{V}_jV_i\right) + \frac{\Delta t^2}{8}\mathcal{V}_k\mathcal{V}_j\p_k\p_jV_i \\ &+ \frac{\Delta t^2}{4}\mathcal{V}_j\p_j\left(V_i\right)\p_k\left(\mathcal{V}_k\right) + \order(\Delta t).
\end{split}\end{equation}
All but the second term vanish in expectation, leaving us with
\begin{equation}\label{eqn:expectation_of_simplified_rhs}
    \langle\nu V^m_i\rangle = \frac{\Delta t}{2}\p_j\langle \mathcal{V}_jV_i\rangle + \order(\Delta t).
\end{equation}
Note that if we are calculating the divergence using RFD instead, \cref{eqn:consistency_full_rhs,eqn:consistency_simplified_rhs} only hold in expectation and are obtained by exploiting the independence of $\v{W}$ and $\tilde{\v{W}}$ to separate terms. Nevertheless, we obtain \cref{eqn:expectation_of_simplified_rhs}.

Now, recalling the definitions of $\v{\mathcal{V}}$ and $\v{V}$ we have
\begin{equation}\begin{split}
    \langle \mathcal{V}_jV_i\rangle = \frac{2k_BT}{\Delta t}N_{ik}K^\top_{kl}M^{-1}_{lm}M^{1/2}_{mn}\left(K^\top K\right)^{-1}_{jq}K^\top_{qr}M^{1/2}_{rs}\langle W_nW_s\rangle.
\end{split}\end{equation}
Using $\langle W_nW_s\rangle = \delta_{ns}$ it is easy to verify that
\begin{equation}
    \langle \mathcal{V}_jV_i\rangle = \frac{2k_BT}{\Delta t}N_{ij}.
\end{equation}
Substituting this into \cref{eqn:expectation_of_simplified_rhs} and multiplying through by $\Delta t$ reveals that \cref{eqn:first_moment} is satisfied to first-order. To show the same for \cref{eqn:second_moment}, recall \cref{eqn:consistency_simplified_rhs} and observe that
\begin{equation}\label{eqn:second_moment_expansion}\begin{split}
    \Delta x_i\Delta x_a &= \Delta t^2 \nu^2 V^m_iV^m_a \\ &= \Delta t^2 V_iV_a + \frac{\Delta t^3}{2}\left[\p_j\left(V_i\mathcal{V}_jV_a\right) + V_iV_a\p_j\mathcal{V}_j\right] + \order(\Delta t^2).
\end{split}\end{equation}
The second term vanishes in expectation and we have
\begin{equation}
    \langle\Delta x_i\Delta x_a\rangle = \Delta t^2\langle V_i V_a\rangle + \order(\Delta t^2).
    \label{eq:second_moment_ViVa}
\end{equation}
If we are using RFD to calculate the divergence, then the expansion in \cref{eqn:second_moment_expansion} only holds in expectation but we obtain \cref{eq:second_moment_ViVa} nonetheless.

Again recalling the definition of $\v{V}$ from \cref{sec:gDC_alg} we find that
\begin{equation}
    \Delta t^2\langle V_iV_a\rangle = 2k_BT\Delta tN_{ik}K^\top_{kl}M^{-1}_{lm}M^{1/2}_{mn}N_{ab}K^\top_{bc}M^{-1}_{cd}M^{1/2}_{de}\langle W_nW_e\rangle.
\end{equation}
In a similar way to before, substituting $\langle W_nW_e\rangle = \delta_{ne}$ and summing over repeated indices yields
\begin{equation}
    \Delta t^2\langle V_iV_a\rangle = 2k_BT\Delta tN_{ia},
\end{equation}
giving us the desired result.

\pagebreak
\addcontentsline{toc}{section}{References}
\bibliography{ref.bib}

\end{document}

%% file: figs/drawing.pdf_tex
\begingroup%
  \makeatletter%
  \providecommand\color[2][]{%
    \errmessage{(Inkscape) Color is used for the text in Inkscape, but the package 'color.sty' is not loaded}%
    \renewcommand\color[2][]{}%
  }%
  \providecommand\transparent[1]{%
    \errmessage{(Inkscape) Transparency is used (non-zero) for the text in Inkscape, but the package 'transparent.sty' is not loaded}%
    \renewcommand\transparent[1]{}%
  }%
  \providecommand\rotatebox[2]{#2}%
  \newcommand*\fsize{\dimexpr\f@size pt\relax}%
  \newcommand*\lineheight[1]{\fontsize{\fsize}{#1\fsize}\selectfont}%
  \ifx\svgwidth\undefined%
    \setlength{\unitlength}{583.93700787bp}%
    \ifx\svgscale\undefined%
      \relax%
    \else%
      \setlength{\unitlength}{\unitlength * \real{\svgscale}}%
    \fi%
  \else%
    \setlength{\unitlength}{\svgwidth}%
  \fi%
  \global\let\svgwidth\undefined%
  \global\let\svgscale\undefined%
  \makeatother%
  \begin{picture}(1,0.48998249)%
    \lineheight{1}%
    \setlength\tabcolsep{0pt}%
    \put(0,0){\includegraphics[width=\unitlength,page=1]{drawing.pdf}}%
    \put(0.43242939,0.24152291){\color[rgb]{0,0,0}\makebox(0,0)[lt]{\lineheight{1.25}\smash{\begin{tabular}[t]{l}$\v{Y}_p$\end{tabular}}}}%
    \put(0,0){\includegraphics[width=\unitlength,page=2]{drawing.pdf}}%
    \put(0.57438096,0.24160617){\color[rgb]{0,0,0}\makebox(0,0)[lt]{\lineheight{1.25}\smash{\begin{tabular}[t]{l}$\v{b} = \v{B}(0)$\end{tabular}}}}%
    \put(0.56853084,0.32838416){\color[rgb]{0,0,0}\makebox(0,0)[lt]{\lineheight{1.25}\smash{\begin{tabular}[t]{l}$\v{B}(t) = {\color{blue}\t{R}(\q{q}_p(t))} \v{b}$\end{tabular}}}}%
    \put(0,0){\includegraphics[width=\unitlength,page=3]{drawing.pdf}}%
  \end{picture}%
\endgroup%

%% file: figs/drawing_discrete.pdf_tex
\begingroup%
  \makeatletter%
  \providecommand\color[2][]{%
    \errmessage{(Inkscape) Color is used for the text in Inkscape, but the package 'color.sty' is not loaded}%
    \renewcommand\color[2][]{}%
  }%
  \providecommand\transparent[1]{%
    \errmessage{(Inkscape) Transparency is used (non-zero) for the text in Inkscape, but the package 'transparent.sty' is not loaded}%
    \renewcommand\transparent[1]{}%
  }%
  \providecommand\rotatebox[2]{#2}%
  \newcommand*\fsize{\dimexpr\f@size pt\relax}%
  \newcommand*\lineheight[1]{\fontsize{\fsize}{#1\fsize}\selectfont}%
  \ifx\svgwidth\undefined%
    \setlength{\unitlength}{340.15748031bp}%
    \ifx\svgscale\undefined%
      \relax%
    \else%
      \setlength{\unitlength}{\unitlength * \real{\svgscale}}%
    \fi%
  \else%
    \setlength{\unitlength}{\svgwidth}%
  \fi%
  \global\let\svgwidth\undefined%
  \global\let\svgscale\undefined%
  \makeatother%
  \begin{picture}(1,1.21957016)%
    \lineheight{1}%
    \setlength\tabcolsep{0pt}%
    \put(0,0){\includegraphics[width=\unitlength,page=1]{drawing_discrete.pdf}}%
    \put(0.49535724,0.57186971){\color[rgb]{0,0,0}\makebox(0,0)[lt]{\lineheight{1.25}\smash{\begin{tabular}[t]{l}$\v{Y}_p$\end{tabular}}}}%
    \put(0,0){\includegraphics[width=\unitlength,page=2]{drawing_discrete.pdf}}%
    \put(0.36491245,0.97373823){\color[rgb]{1,0,0}\makebox(0,0)[lt]{\lineheight{1.25}\smash{\begin{tabular}[t]{l}$\v{r}_i$\end{tabular}}}}%
    \put(0,0){\includegraphics[width=\unitlength,page=3]{drawing_discrete.pdf}}%
    \put(0.55253403,0.48096595){\color[rgb]{0,0,1}\makebox(0,0)[lt]{\lineheight{1.25}\smash{\begin{tabular}[t]{l}$\v{\Omega}_p$\end{tabular}}}}%
    \put(0.35174649,0.66476245){\color[rgb]{0,0,0}\makebox(0,0)[lt]{\lineheight{1.25}\smash{\begin{tabular}[t]{l}$\dot{\v{Y}}_p$\end{tabular}}}}%
    \put(0,0){\includegraphics[width=\unitlength,page=4]{drawing_discrete.pdf}}%
    \put(0.15072584,1.151702){\color[rgb]{1,0,0}\makebox(0,0)[lt]{\lineheight{1.25}\smash{\begin{tabular}[t]{l}$\fd{\v{r}_i}{t} = \dot{\v{Y}}_p + \v{\Omega}_p\times\left(\v{r}_i - \v{Y}_p\right)$\end{tabular}}}}%
  \end{picture}%
\endgroup%

%% file: figs/hydrophobic_radial_distributions_ignoring_largest_f0_zoomed_in_with_arrow.pdf_tex
\begingroup%
  \makeatletter%
  \providecommand\color[2][]{%
    \errmessage{(Inkscape) Color is used for the text in Inkscape, but the package 'color.sty' is not loaded}%
    \renewcommand\color[2][]{}%
  }%
  \providecommand\transparent[1]{%
    \errmessage{(Inkscape) Transparency is used (non-zero) for the text in Inkscape, but the package 'transparent.sty' is not loaded}%
    \renewcommand\transparent[1]{}%
  }%
  \providecommand\rotatebox[2]{#2}%
  \newcommand*\fsize{\dimexpr\f@size pt\relax}%
  \newcommand*\lineheight[1]{\fontsize{\fsize}{#1\fsize}\selectfont}%
  \ifx\svgwidth\undefined%
    \setlength{\unitlength}{516bp}%
    \ifx\svgscale\undefined%
      \relax%
    \else%
      \setlength{\unitlength}{\unitlength * \real{\svgscale}}%
    \fi%
  \else%
    \setlength{\unitlength}{\svgwidth}%
  \fi%
  \global\let\svgwidth\undefined%
  \global\let\svgscale\undefined%
  \makeatother%
  \begin{picture}(1,0.76356589)%
    \lineheight{1}%
    \setlength\tabcolsep{0pt}%
    \put(0,0){\includegraphics[width=\unitlength,page=1]{hydrophobic_radial_distributions_ignoring_largest_f0_zoomed_in_with_arrow.pdf}}%
    \put(0.56661785,0.32139899){\color[rgb]{0,0,1}\makebox(0,0)[lt]{\lineheight{1.25}\smash{\begin{tabular}[t]{l}Increasing $f_0$\end{tabular}}}}%
  \end{picture}%
\endgroup%

%% file: main.bbl
\begin{thebibliography}{10}
\expandafter\ifx\csname url\endcsname\relax
  \def\url#1{\texttt{#1}}\fi
\expandafter\ifx\csname urlprefix\endcsname\relax\def\urlprefix{URL }\fi
\expandafter\ifx\csname href\endcsname\relax
  \def\href#1#2{#2} \def\path#1{#1}\fi

\bibitem{Russel1981}
W.~B. Russel, Brownian motion of small particles suspended in liquids, Annual
  Review of Fluid Mechanics 13~(1) (1981) 425--455.
\newblock \href {http://dx.doi.org/10.1146/annurev.fl.13.010181.002233}
  {\path{doi:10.1146/annurev.fl.13.010181.002233}}.

\bibitem{russel1991colloidal}
W.~B. Russel, W.~Russel, D.~A. Saville, W.~R. Schowalter, Colloidal
  dispersions, Cambridge university press, 1991.

\bibitem{graham2018microhydrodynamics}
M.~D. Graham, Microhydrodynamics, Brownian motion, and complex fluids, Vol.~58,
  Cambridge University Press, 2018.

\bibitem{doi_theory_1986}
M.~Doi, S.~F. Edwards, The {Theory} of {Polymer} {Dynamics}, International
  {Series} of {Monographs} on {Physics}, Oxford Science Publications, 1986.

\bibitem{larson1999structure}
R.~G. Larson, The structure and rheology of complex fluids, Vol. 150, Oxford
  university press New York, 1999.

\bibitem{Batchelor1977}
G.~K. Batchelor, The effect of {B}rownian motion on the bulk stress in a
  suspension of spherical particles, Journal of Fluid Mechanics 83 (1977)
  97--117.
\newblock \href {http://dx.doi.org/10.1017/S0022112077001062}
  {\path{doi:10.1017/S0022112077001062}}.

\bibitem{Bossis1989}
G.~Bossis, J.~F. Brady, The rheology of brownian suspensions, The Journal of
  chemical physics 91~(3) (1989) 1866--1874.

\bibitem{Foss2000}
D.~R. Foss, J.~F. Brady, Structure, diffusion and rheology of {B}rownian
  suspensions by {S}tokesian dynamics simulation, Journal of Fluid Mechanics
  407 (2000) 167--200.
\newblock \href {http://dx.doi.org/10.1017/S0022112099007557}
  {\path{doi:10.1017/S0022112099007557}}.

\bibitem{bressloff2013stochastic}
P.~C. Bressloff, J.~M. Newby, Stochastic models of intracellular transport,
  Reviews of Modern Physics 85~(1) (2013) 135.

\bibitem{ermak_brownian_1978}
D.~L. Ermak, J.~A. McCammon,
  \href{https://aip.scitation.org/doi/10.1063/1.436761}{Brownian dynamics with
  hydrodynamic interactions}, The Journal of Chemical Physics 69~(4) (1978)
  1352--1360.
\newblock \href {http://dx.doi.org/10.1063/1.436761}
  {\path{doi:10.1063/1.436761}}.
\newline\urlprefix\url{https://aip.scitation.org/doi/10.1063/1.436761}

\bibitem{pavliotis2014stochastic}
G.~A. Pavliotis, Stochastic processes and applications: diffusion processes,
  the Fokker-Planck and Langevin equations, Vol.~60, Springer, 2014.

\bibitem{Kim1991}
S.~Kim, S.~J. Karrila, Microhydrodynamics: principles and selected
  applications, Courier Dover Publications, 1991.

\bibitem{Happel2012}
J.~Happel, H.~Brenner, Low Reynolds number hydrodynamics: with special
  applications to particulate media, Vol.~1, Springer Science \& Business
  Media, 2012.

\bibitem{kubo_fluctuation-dissipation_1966}
R.~Kubo, \href{https://doi.org/10.1088%2F0034-4885%2F29%2F1%2F306}{The
  fluctuation-dissipation theorem}, Reports on Progress in Physics 29~(1)
  (1966) 255--284.
\newblock \href {http://dx.doi.org/10.1088/0034-4885/29/1/306}
  {\path{doi:10.1088/0034-4885/29/1/306}}.
\newline\urlprefix\url{https://doi.org/10.1088%2F0034-4885%2F29%2F1%2F306}

\bibitem{Brady1988}
J.~F. Brady, G.~Bossis, Stokesian dynamics, Annual Review of Fluid Mechanics
  20~(1) (1988) 111--157.
\newblock \href {http://dx.doi.org/10.1146/annurev.fl.20.010188.000551}
  {\path{doi:10.1146/annurev.fl.20.010188.000551}}.

\bibitem{sierou2001accelerated}
A.~Sierou, J.~F. Brady, Accelerated stokesian dynamics simulations, Journal of
  fluid mechanics 448 (2001) 115--146.

\bibitem{Banchio2003}
A.~J. Banchio, J.~F. Brady, Accelerated stokesian dynamics: Brownian motion,
  The Journal of Chemical Physics 118~(22) (2003) 10323--10332.
\newblock \href {http://dx.doi.org/10.1063/1.1571819}
  {\path{doi:10.1063/1.1571819}}.

\bibitem{greengard1987fast}
L.~Greengard, V.~Rokhlin, A fast algorithm for particle simulations, Journal of
  computational physics 73~(2) (1987) 325--348.

\bibitem{tornberg2008fast}
A.-K. Tornberg, L.~Greengard, A fast multipole method for the three-dimensional
  stokes equations, Journal of Computational Physics 227~(3) (2008) 1613--1619.

\bibitem{ying2004kernel}
L.~Ying, G.~Biros, D.~Zorin, A kernel-independent adaptive fast multipole
  algorithm in two and three dimensions, Journal of Computational Physics
  196~(2) (2004) 591--626.

\bibitem{lindbo2011spectral}
D.~Lindbo, A.-K. Tornberg, Spectral accuracy in fast ewald-based methods for
  particle simulations, Journal of Computational Physics 230~(24) (2011)
  8744--8761.

\bibitem{Fiore2017}
A.~M. Fiore, F.~Balboa~Usabiaga, A.~Donev, J.~W. Swan, Rapid sampling of
  stochastic displacements in brownian dynamics simulations, The Journal of
  chemical physics 146~(12) (2017) 124116.

\bibitem{Fiore2018}
A.~M. Fiore, J.~W. Swan, Rapid sampling of stochastic displacements in brownian
  dynamics simulations with stresslet constraints, The Journal of chemical
  physics 148~(4) (2018) 044114.

\bibitem{wang2016spectral}
M.~Wang, J.~F. Brady, Spectral ewald acceleration of stokesian dynamics for
  polydisperse suspensions, Journal of Computational Physics 306 (2016)
  443--477.

\bibitem{fiore2019fast}
A.~M. Fiore, J.~W. Swan, Fast stokesian dynamics, Journal of Fluid Mechanics
  878 (2019) 544--597.

\bibitem{Peskin2002}
C.~Peskin, {The immersed boundary method}, {Acta Numerica} {11} ({2002})
  {479--517}.

\bibitem{maxey_localized_2001-1}
M.~R. Maxey, B.~K. Patel, Localized force representations for particles
  sedimenting in {Stokes} ¯ow, International Journal of Multiphase Flow (2001)
  24.

\bibitem{lomholt_force-coupling_2003-1}
S.~Lomholt, M.~R. Maxey,
  \href{http://linkinghub.elsevier.com/retrieve/pii/S0021999102000219}{Force-coupling
  method for particulate two-phase flow: {Stokes} flow}, Journal of
  Computational Physics 184~(2) (2003) 381--405.
\newblock \href {http://dx.doi.org/10.1016/S0021-9991(02)00021-9}
  {\path{doi:10.1016/S0021-9991(02)00021-9}}.
\newline\urlprefix\url{http://linkinghub.elsevier.com/retrieve/pii/S0021999102000219}

\bibitem{Fixman1986}
M.~Fixman, Construction of {L}angevin forces in the simulation of hydrodynamic
  interaction, Macromolecules 19~(4) (1986) 1204--1207.
\newblock \href {http://dx.doi.org/10.1021/ma00158a043}
  {\path{doi:10.1021/ma00158a043}}.

\bibitem{jendrejack2000hydrodynamic}
R.~M. Jendrejack, M.~D. Graham, J.~J. de~Pablo, Hydrodynamic interactions in
  long chain polymers: Application of the chebyshev polynomial approximation in
  stochastic simulations, The Journal of Chemical Physics 113~(7) (2000)
  2894--2900.

\bibitem{chow_preconditioned_2014}
E.~Chow, Y.~Saad,
  \href{http://epubs.siam.org/doi/10.1137/130920587}{Preconditioned {Krylov}
  {Subspace} {Methods} for {Sampling} {Multivariate} {Gaussian}
  {Distributions}}, SIAM Journal on Scientific Computing 36~(2) (2014)
  A588--A608.
\newblock \href {http://dx.doi.org/10.1137/130920587}
  {\path{doi:10.1137/130920587}}.
\newline\urlprefix\url{http://epubs.siam.org/doi/10.1137/130920587}

\bibitem{Landau1959}
L.~Landau, E.~Lifshitz, Fluid Mechanics, Pergamon Press, 1959.

\bibitem{Ladd1993}
A.~J.~C. Ladd, Short-time motion of colloidal particles: Numerical simulation
  via a fluctuating lattice-{B}oltzmann equation, Phys. Rev. Lett. 70 (1993)
  1339--1342.
\newblock \href {http://dx.doi.org/10.1103/PhysRevLett.70.1339}
  {\path{doi:10.1103/PhysRevLett.70.1339}}.

\bibitem{Sharma2004}
N.~Sharma, N.~A. Patankar, Direct numerical simulation of the {B}rownian motion
  of particles by using fluctuating hydrodynamic equations, Journal of
  Computational Physics 201~(2) (2004) 466 -- 486.
\newblock \href {http://dx.doi.org/10.1016/j.jcp.2004.06.002}
  {\path{doi:10.1016/j.jcp.2004.06.002}}.

\bibitem{de2016finite}
M.~De~Corato, J.~Slot, M.~H{\"u}tter, G.~D'Avino, P.~L. Maffettone, M.~A.
  Hulsen, Finite element formulation of fluctuating hydrodynamics for fluids
  filled with rigid particles using boundary fitted meshes, Journal of
  Computational Physics 316 (2016) 632--651.

\bibitem{Delong2014}
S.~Delong, F.~B. Usabiaga, R.~Delgado-Buscalioni, B.~E. Griffith, A.~Donev,
  Brownian dynamics without green's functions, The Journal of chemical physics
  140~(13) (2014) 134110.

\bibitem{atzberger2007stochastic}
P.~J. Atzberger, P.~R. Kramer, C.~S. Peskin, A stochastic immersed boundary
  method for fluid-structure dynamics at microscopic length scales, Journal of
  Computational Physics 224~(2) (2007) 1255--1292.

\bibitem{kramer2008foundations}
P.~R. Kramer, C.~S. Peskin, P.~J. Atzberger, On the foundations of the
  stochastic immersed boundary method, Computer Methods in Applied Mechanics
  and Engineering 197~(25-28) (2008) 2232--2249.

\bibitem{keaveny_fluctuating_2014-1}
E.~E. Keaveny,
  \href{http://www.sciencedirect.com/science/article/pii/S0021999114001867}{Fluctuating
  force-coupling method for simulations of colloidal suspensions}, Journal of
  Computational Physics 269 (2014) 61--79.
\newblock \href {http://dx.doi.org/10.1016/j.jcp.2014.03.013}
  {\path{doi:10.1016/j.jcp.2014.03.013}}.
\newline\urlprefix\url{http://www.sciencedirect.com/science/article/pii/S0021999114001867}

\bibitem{delmotte_simulating_2015}
B.~Delmotte, E.~E. Keaveny,
  \href{http://aip.scitation.org/doi/10.1063/1.4938173}{Simulating {Brownian}
  suspensions with fluctuating hydrodynamics}, The Journal of Chemical Physics
  143~(24) (2015) 244109.
\newblock \href {http://dx.doi.org/10.1063/1.4938173}
  {\path{doi:10.1063/1.4938173}}.
\newline\urlprefix\url{http://aip.scitation.org/doi/10.1063/1.4938173}

\bibitem{bao2018fluctuating}
Y.~Bao, M.~Rachh, E.~E. Keaveny, L.~Greengard, A.~Donev, A fluctuating boundary
  integral method for brownian suspensions, Journal of Computational Physics
  374 (2018) 1094--1119.

\bibitem{Fixman1978}
M.~Fixman, Simulation of polymer dynamics. {I}. {G}eneral theory, The Journal
  of Chemical Physics 69~(4) (1978) 1527--1537.
\newblock \href {http://dx.doi.org/10.1063/1.436725}
  {\path{doi:10.1063/1.436725}}.

\bibitem{Grassia1995}
P.~S. Grassia, E.~J. Hinch, L.~C. Nitsche, Computer simulations of {B}rownian
  motion of complex systems, Journal of Fluid Mechanics 282 (1995) 373--403.
\newblock \href {http://dx.doi.org/10.1017/S0022112095000176}
  {\path{doi:10.1017/S0022112095000176}}.

\bibitem{sprinkle_large_2017}
B.~Sprinkle, F.~B. Usabiaga, N.~A. Patankar, A.~Donev,
  \href{http://arxiv.org/abs/1709.02410}{Large {Scale} {Brownian} {Dynamics} of
  {Confined} {Suspensions} of {Rigid} {Particles}}, The Journal of Chemical
  Physics 147~(24) (2017) 244103, arXiv: 1709.02410.
\newblock \href {http://dx.doi.org/10.1063/1.5003833}
  {\path{doi:10.1063/1.5003833}}.
\newline\urlprefix\url{http://arxiv.org/abs/1709.02410}

\bibitem{Sprinkle2019}
B.~Sprinkle, A.~Donev, A.~P.~S. Bhalla, N.~Patankar, Brownian dynamics of fully
  confined suspensions of rigid particles without green’s functions, The
  Journal of chemical physics 150~(16) (2019) 164116.

\bibitem{Pozrikidis}
C.~Pozrikidis, Boundary integral and singularity methods for linearized viscous
  flow, Cambridge University Press, 1992.

\bibitem{Balboa2017}
F.~Balboa~Usabiaga, B.~Delmotte, A.~Donev, Brownian dynamics of confined
  suspensions of active microrollers, The Journal of chemical physics 146~(13)
  (2017) 134104.

\bibitem{usabiaga_hydrodynamics_2016}
F.~B. Usabiaga, B.~Kallemov, B.~Delmotte, A.~P.~S. Bhalla, B.~E. Griffith,
  A.~Donev, \href{http://arxiv.org/abs/1602.02170}{Hydrodynamics of
  {Suspensions} of {Passive} and {Active} {Rigid} {Particles}: {A} {Rigid}
  {Multiblob} {Approach}}, Communications in Applied Mathematics and
  Computational Science 11~(2) (2016) 217--296, arXiv: 1602.02170.
\newblock \href {http://dx.doi.org/10.2140/camcos.2016.11.217}
  {\path{doi:10.2140/camcos.2016.11.217}}.
\newline\urlprefix\url{http://arxiv.org/abs/1602.02170}

\bibitem{wajnryb_generalization_2013}
E.~Wajnryb, K.~A. Mizerski, P.~J. Zuk, P.~Szymczak,
  \href{http://arxiv.org/abs/1307.7312}{Generalization of the
  {Rotne}-{Prager}-{Yamakawa} mobility and shear disturbance tensors}, Journal
  of Fluid Mechanics 731, arXiv: 1307.7312.
\newblock \href {http://dx.doi.org/10.1017/jfm.2013.402}
  {\path{doi:10.1017/jfm.2013.402}}.
\newline\urlprefix\url{http://arxiv.org/abs/1307.7312}

\bibitem{zuk_rotneprageryamakawa_2014}
P.~J. Zuk, E.~Wajnryb, K.~A. Mizerski, P.~Szymczak,
  \href{http://www.journals.cambridge.org/abstract_S002211201300668X}{Rotne–{Prager}–{Yamakawa}
  approximation for different-sized particles in application to macromolecular
  bead models}, Journal of Fluid Mechanics 741.
\newblock \href {http://dx.doi.org/10.1017/jfm.2013.668}
  {\path{doi:10.1017/jfm.2013.668}}.
\newline\urlprefix\url{http://www.journals.cambridge.org/abstract_S002211201300668X}

\bibitem{swan_simulation_2007}
J.~W. Swan, J.~F. Brady,
  \href{http://aip.scitation.org/doi/10.1063/1.2803837}{Simulation of
  hydrodynamically interacting particles near a no-slip boundary}, Physics of
  Fluids 19~(11) (2007) 113306.
\newblock \href {http://dx.doi.org/10.1063/1.2803837}
  {\path{doi:10.1063/1.2803837}}.
\newline\urlprefix\url{http://aip.scitation.org/doi/10.1063/1.2803837}

\bibitem{Liang2013}
Z.~Liang, Z.~Gimbutas, L.~Greengard, J.~Huang, S.~Jiang, A fast multipole
  method for the rotne--prager--yamakawa tensor and its applications, Journal
  of Computational Physics 234 (2013) 133--139.

\bibitem{keaveny_fluctuating_2014}
E.~E. Keaveny,
  \href{http://linkinghub.elsevier.com/retrieve/pii/S0021999114001867}{Fluctuating
  force-coupling method for simulations of colloidal suspensions}, Journal of
  Computational Physics 269 (2014) 61--79.
\newblock \href {http://dx.doi.org/10.1016/j.jcp.2014.03.013}
  {\path{doi:10.1016/j.jcp.2014.03.013}}.
\newline\urlprefix\url{http://linkinghub.elsevier.com/retrieve/pii/S0021999114001867}

\bibitem{maxey_localized_2001}
M.~R. Maxey, B.~K. Patel, Localized force representations for particles
  sedimenting in {Stokes} fow, Int. J. Multiphase Flow 27 (2001) 1603--1626.

\bibitem{delmotte_general_2015}
B.~Delmotte, E.~Climent, F.~Plouraboué,
  \href{http://dx.doi.org/10.1016/j.jcp.2015.01.026}{A general formulation of
  {Bead} {Models} applied to flexible fibers and active filaments at low
  {Reynolds} number}, J. Comput. Phys. 286 (2015) 14--37.
\newblock \href {http://dx.doi.org/10.1016/j.jcp.2015.01.026}
  {\path{doi:10.1016/j.jcp.2015.01.026}}.
\newline\urlprefix\url{http://dx.doi.org/10.1016/j.jcp.2015.01.026}

\bibitem{Atzberger2007}
P.~J. Atzberger, P.~R. Kramer, C.~S. Peskin, A stochastic immersed boundary
  method for fluid-structure dynamics at microscopic length scales, Journal of
  Computational Physics 224~(2) (2007) 1255 -- 1292.
\newblock \href {http://dx.doi.org/10.1016/j.jcp.2006.11.015}
  {\path{doi:10.1016/j.jcp.2006.11.015}}.

\bibitem{Atzberger2011}
P.~J. Atzberger, Stochastic eulerian lagrangian methods for fluid�structure
  interactions with thermal fluctuations, Journal of Computational Physics
  230~(8) (2011) 2821 -- 2837.
\newblock \href {http://dx.doi.org/10.1016/j.jcp.2010.12.028}
  {\path{doi:10.1016/j.jcp.2010.12.028}}.

\bibitem{Bourrianne2020}
P.~Bourrianne, V.~Niggel, G.~Polly, T.~Divoux, G.~H. McKinley, Unifying
  disparate experimental views on shear-thickening suspensions (2020).
\newblock \href {http://arxiv.org/abs/2001.02290} {\path{arXiv:2001.02290}}.

\bibitem{driscoll_unstable_2017}
M.~Driscoll, B.~Delmotte, M.~Youssef, S.~Sacanna, A.~Donev, P.~Chaikin,
  \href{http://arxiv.org/abs/1609.08673}{Unstable fronts and stable "critters"
  formed by microrollers}, Nature Physics 13~(4) (2017) 375--379, arXiv:
  1609.08673.
\newblock \href {http://dx.doi.org/10.1038/nphys3970}
  {\path{doi:10.1038/nphys3970}}.
\newline\urlprefix\url{http://arxiv.org/abs/1609.08673}

\bibitem{delong_brownian_2015}
S.~Delong, F.~Balboa~Usabiaga, A.~Donev,
  \href{http://aip.scitation.org/doi/10.1063/1.4932062}{Brownian dynamics of
  confined rigid bodies}, The Journal of Chemical Physics 143~(14) (2015)
  144107.
\newblock \href {http://dx.doi.org/10.1063/1.4932062}
  {\path{doi:10.1063/1.4932062}}.
\newline\urlprefix\url{http://aip.scitation.org/doi/10.1063/1.4932062}

\bibitem{Chakrabarty2013}
A.~Chakrabarty, A.~Konya, F.~Wang, J.~V. Selinger, K.~Sun, Q.-H. Wei, Brownian
  motion of boomerang colloidal particles, Physical review letters 111~(16)
  (2013) 160603.

\bibitem{eshraghi_molecular_2018}
M.~Eshraghi, J.~Horbach,
  \href{https://pubs.rsc.org/en/content/articlelanding/2018/sm/c8sm00398j}{Molecular
  dynamics simulation of charged colloids confined between hard walls:
  pre-melting and pre-freezing across the {BCC}–fluid coexistence}, Soft
  Matter 14~(20) (2018) 4141--4149.
\newblock \href {http://dx.doi.org/10.1039/C8SM00398J}
  {\path{doi:10.1039/C8SM00398J}}.
\newline\urlprefix\url{https://pubs.rsc.org/en/content/articlelanding/2018/sm/c8sm00398j}

\bibitem{snook_monte_1978}
I.~K. Snook, D.~Henderson,
  \href{https://aip.scitation.org/doi/10.1063/1.436036}{Monte {Carlo} study of
  a hard‐sphere fluid near a hard wall}, The Journal of Chemical Physics
  68~(5) (1978) 2134--2139.
\newblock \href {http://dx.doi.org/10.1063/1.436036}
  {\path{doi:10.1063/1.436036}}.
\newline\urlprefix\url{https://aip.scitation.org/doi/10.1063/1.436036}

\bibitem{mittal_does_2007}
J.~Mittal, J.~R. Errington, T.~M. Truskett,
  \href{https://aip-scitation-org.iclibezp1.cc.ic.ac.uk/doi/10.1063/1.2748045}{Does
  confining the hard-sphere fluid between hard walls change its average
  properties?}, The Journal of Chemical Physics 126~(24) (2007) 244708.
\newblock \href {http://dx.doi.org/10.1063/1.2748045}
  {\path{doi:10.1063/1.2748045}}.
\newline\urlprefix\url{https://aip-scitation-org.iclibezp1.cc.ic.ac.uk/doi/10.1063/1.2748045}

\bibitem{deb_hard_2011}
D.~Deb, A.~Winkler, M.~H. Yamani, M.~Oettel, P.~Virnau, K.~Binder,
  \href{https://aip-scitation-org.iclibezp1.cc.ic.ac.uk/doi/10.1063/1.3593197}{Hard
  sphere fluids at a soft repulsive wall: {A} comparative study using {Monte}
  {Carlo} and density functional methods}, The Journal of Chemical Physics
  134~(21) (2011) 214706.
\newblock \href {http://dx.doi.org/10.1063/1.3593197}
  {\path{doi:10.1063/1.3593197}}.
\newline\urlprefix\url{https://aip-scitation-org.iclibezp1.cc.ic.ac.uk/doi/10.1063/1.3593197}

\bibitem{van_winkle_layering_1988}
D.~H. Van~Winkle, C.~A. Murray,
  \href{https://aip-scitation-org.iclibezp1.cc.ic.ac.uk/doi/10.1063/1.454864}{Layering
  in colloidal fluids near a smooth repulsive wall}, The Journal of Chemical
  Physics 89~(6) (1988) 3885--3891.
\newblock \href {http://dx.doi.org/10.1063/1.454864}
  {\path{doi:10.1063/1.454864}}.
\newline\urlprefix\url{https://aip-scitation-org.iclibezp1.cc.ic.ac.uk/doi/10.1063/1.454864}

\bibitem{bourrianne_unifying_2020}
P.~Bourrianne, V.~Niggel, G.~Polly, T.~Divoux, G.~H. McKinley,
  \href{http://arxiv.org/abs/2001.02290}{Unifying disparate experimental views
  on shear-thickening suspensions}, arXiv:2001.02290 [cond-mat,
  physics:physics]ArXiv: 2001.02290.
\newline\urlprefix\url{http://arxiv.org/abs/2001.02290}

\bibitem{vazquez-quesada_multiblob_2014}
A.~Vázquez-Quesada, F.~Balboa~Usabiaga, R.~Delgado-Buscalioni,
  \href{https://aip.scitation.org/doi/10.1063/1.4901889}{A multiblob approach
  to colloidal hydrodynamics with inherent lubrication}, The Journal of
  Chemical Physics 141~(20) (2014) 204102.
\newblock \href {http://dx.doi.org/10.1063/1.4901889}
  {\path{doi:10.1063/1.4901889}}.
\newline\urlprefix\url{https://aip.scitation.org/doi/10.1063/1.4901889}

\bibitem{iserles_lie-group_2000}
A.~Iserles, H.~Z. Munthe-Kaas, S.~P. Nørsett, A.~Zanna,
  \href{https://www.cambridge.org/core/journals/acta-numerica/article/liegroup-methods/856125FF1EAF7762DEF6E37EEBA9CA5F}{Lie-group
  methods}, Acta Numerica 9 (2000) 215--365.
\newline\urlprefix\url{https://www.cambridge.org/core/journals/acta-numerica/article/liegroup-methods/856125FF1EAF7762DEF6E37EEBA9CA5F}

\bibitem{faltinsen_multistep_2001}
S.~Faltinsen, A.~Marthinsen, H.~Z. Munthe-Kaas,
  \href{http://www.sciencedirect.com/science/article/pii/S0168927401001039}{Multistep
  methods integrating ordinary differential equations on manifolds}, Applied
  Numerical Mathematics 39~(3) (2001) 349--365.
\newblock \href {http://dx.doi.org/10.1016/S0168-9274(01)00103-9}
  {\path{doi:10.1016/S0168-9274(01)00103-9}}.
\newline\urlprefix\url{http://www.sciencedirect.com/science/article/pii/S0168927401001039}

\bibitem{Malham_2008}
S.~J.~A. Malham, A.~Wiese,
  \href{http://dx.doi.org/10.1137/060666743}{Stochastic lie group integrators},
  SIAM Journal on Scientific Computing 30~(2) (2008) 597–617.
\newblock \href {http://dx.doi.org/10.1137/060666743}
  {\path{doi:10.1137/060666743}}.
\newline\urlprefix\url{http://dx.doi.org/10.1137/060666743}

\end{thebibliography}
